\documentclass[floatfix, noshowpacs, preprintnumbers, twocolumn, amsmath, amssymb, aps, prl, superscriptaddress]{revtex4}

\pdfoutput=1

\usepackage{graphicx, xcolor}
\usepackage[caption=false]{subfig}
\usepackage{soul}
\usepackage{slashed}
\usepackage{bm} \usepackage{graphicx} \usepackage{amsmath}
\usepackage{amsthm,stmaryrd}
\usepackage{amssymb} \usepackage{physics} \usepackage{txfonts}
\usepackage{color}
\usepackage{comment}
\usepackage{hyperref}

\begin{document}

\title{Generating Haar-uniform Randomness using Stochastic Quantum Walks \\ on a Photonic Chip}

\author{Hao Tang}
\affiliation{Center for Integrated Quantum Information Technologies (IQIT), School of Physics and Astronomy and State Key Laboratory of Advanced Optical Communication Systems and Networks, Shanghai Jiao Tong University, Shanghai 200240, China}
\affiliation{CAS Center for Excellence and Synergetic Innovation Center in Quantum Information and Quantum Physics, University of Science and Technology of China, Hefei, Anhui 230026, China}

\author{Leonardo Banchi}
\affiliation{Department of Physics and Astronomy, University of Florence, via G. Sansone 1, I-50019 Sesto Fiorentino (FI), Italy}
\affiliation{INFN Sezione di Firenze, via G. Sansone 1, I-50019 Sesto Fiorentino (FI), Italy}

\author{Tian-Yu Wang} 
\affiliation{Center for Integrated Quantum Information Technologies (IQIT), School of Physics and Astronomy and State Key Laboratory of Advanced Optical Communication Systems and Networks, Shanghai Jiao Tong University, Shanghai 200240, China}
\affiliation{CAS Center for Excellence and Synergetic Innovation Center in Quantum Information and Quantum Physics, University of Science and Technology of China, Hefei, Anhui 230026, China}

\author{Xiao-Wen Shang} 
\affiliation{Center for Integrated Quantum Information Technologies (IQIT), School of Physics and Astronomy and State Key Laboratory of Advanced Optical Communication Systems and Networks, Shanghai Jiao Tong University, Shanghai 200240, China}
\affiliation{CAS Center for Excellence and Synergetic Innovation Center in Quantum Information and Quantum Physics, University of Science and Technology of China, Hefei, Anhui 230026, China}

\author{Xi Tan} 
\affiliation{Center for Integrated Quantum Information Technologies (IQIT), School of Physics and Astronomy and State Key Laboratory of Advanced Optical Communication Systems and Networks, Shanghai Jiao Tong University, Shanghai 200240, China}
\affiliation{CAS Center for Excellence and Synergetic Innovation Center in Quantum Information and Quantum Physics, University of Science and Technology of China, Hefei, Anhui 230026, China}

\author{Wen-Hao Zhou} 
\affiliation{Center for Integrated Quantum Information Technologies (IQIT), School of Physics and Astronomy and State Key Laboratory of Advanced Optical Communication Systems and Networks, Shanghai Jiao Tong University, Shanghai 200240, China}
\affiliation{CAS Center for Excellence and Synergetic Innovation Center in Quantum Information and Quantum Physics, University of Science and Technology of China, Hefei, Anhui 230026, China}

\author{Zhen Feng} 
\affiliation{Center for Integrated Quantum Information Technologies (IQIT), School of Physics and Astronomy and State Key Laboratory of Advanced Optical Communication Systems and Networks, Shanghai Jiao Tong University, Shanghai 200240, China}
\affiliation{CAS Center for Excellence and Synergetic Innovation Center in Quantum Information and Quantum Physics, University of Science and Technology of China, Hefei, Anhui 230026, China}

\author{Anurag Pal} 
\affiliation{Center for Integrated Quantum Information Technologies (IQIT), School of Physics and Astronomy and State Key Laboratory of Advanced Optical Communication Systems and Networks, Shanghai Jiao Tong University, Shanghai 200240, China}
\affiliation{CAS Center for Excellence and Synergetic Innovation Center in Quantum Information and Quantum Physics, University of Science and Technology of China, Hefei, Anhui 230026, China}

\author{Hang Li} 
\affiliation{Center for Integrated Quantum Information Technologies (IQIT), School of Physics and Astronomy and State Key Laboratory of Advanced Optical Communication Systems and Networks, Shanghai Jiao Tong University, Shanghai 200240, China}
\affiliation{CAS Center for Excellence and Synergetic Innovation Center in Quantum Information and Quantum Physics, University of Science and Technology of China, Hefei, Anhui 230026, China}

\author{Cheng-Qiu Hu} 
\affiliation{Center for Integrated Quantum Information Technologies (IQIT), School of Physics and Astronomy and State Key Laboratory of Advanced Optical Communication Systems and Networks, Shanghai Jiao Tong University, Shanghai 200240, China}
\affiliation{CAS Center for Excellence and Synergetic Innovation Center in Quantum Information and Quantum Physics, University of Science and Technology of China, Hefei, Anhui 230026, China}

\author{M.S. Kim} 
\affiliation{QOLS, Blackett Laboratory, Imperial College London, London SW7 2AZ, UK.}
\affiliation{Korea Institute of Advanced Study, Seoul 02455, South Korea.}

\author{Xian-Min Jin} 
\altaffiliation{xianmin.jin@sjtu.edu.cn} 
\affiliation{Center for Integrated Quantum Information Technologies (IQIT), School of Physics and Astronomy and State Key Laboratory of Advanced Optical Communication Systems and Networks, Shanghai Jiao Tong University, Shanghai 200240, China}
\affiliation{CAS Center for Excellence and Synergetic Innovation Center in Quantum Information and Quantum Physics, University of Science and Technology of China, Hefei, Anhui 230026, China}

\begin{abstract}
As random operations for quantum systems are intensively used in various quantum information tasks, a trustworthy measure of the randomness in quantum operations is highly demanded. The Haar measure of randomness is a useful tool with wide applications such as boson sampling. Recently, a theoretical protocol was proposed to combine quantum control theory and driven stochastic quantum walks to generate Haar-uniform random operations. This opens up a promising route to converting classical randomness to quantum randomness. Here, we implement a two-dimensional stochastic quantum  walk on the integrated photonic chip and demonstrate that the average of all distribution profiles converges to the even distribution when the evolution length increases, suggesting the 1-pad Haar-uniform randomness. We further show that our two-dimensional array outperforms the one-dimensional array of the same number of waveguide for the speed of convergence.  Our work demonstrates a scalable and robust way to generate Haar-uniform randomness that can provide useful building blocks to boost future quantum information techniques.
\end{abstract}

\maketitle

Random operations for quantum systems\cite{Guhr1998} play an important role for a large variety of tasks in quantum information processing. Especially, as various studies on boson sampling\cite{Spring2013, Broome2013, Tillmann2013, Crespi2013, Spagnolo2014, Wang2017} have emerged in recent years to demonstrate quantum computational supremacy\cite{Harrow2017, Wu2018}, the Haar random unitary matrices\cite{Zyczkowski1994} required for these studies have drawn ever increasing attention. The Haar measure of randomness is now investigated as more than a theoretical tool, but also as a useful building block for quantum protocols or algorithms. It has wide applications covering boson sampling\cite{Spring2013, Broome2013, Tillmann2013, Spagnolo2014, Wang2017}, quantum cryptography\cite{Hayden2004,Boykin2003}, quantum process tomography\cite{Bendersky2008}, entanglement generation\cite{Hamma2012}, fidelity estimation\cite{Dankert2009} etc, which, therefore, motivated a series of experimental schemes on implementing random or pseudorandom quantum operations\cite{Harrow2009, Emerson2003, Alexander2016}. So far, these experimental schemes decompose a random unitary matrix either by using a large number of quantum gates\cite{Harrow2009, Emerson2003, Alexander2016}, or using photonic beamsplitters and interferometers\cite{Russell2017, Burgwal2017} via Reck/Clements decomposition method\cite{Reck1994, Clements2016}, both of considerably high complexity in implementation. 

An alternative approach to generate Haar uniform random operations has recently been proposed\cite{Banchi2017} using what we call a \emph{stochastic quantum walk}. The rationale is based on quantum control theory, which allows for a coherent driving of permanently coupled quantum systems via classical control pulses and stochastic pulses. Instead of using quantum circuits or programmable photonic networks with beam splitters and phase shifters, in this alternative approach, random operations can be implemented via permanently coupled photonic waveguides by applying stochastic modulations. This scheme, which effectively implements a stochastic version of a continuous-time quantum walk\cite{Whitfield2010}, could be scalable and beneficial for practical quantum experiments including larger-scale boson sampling. However, up to now, this scheme has never been demonstrated in experiments. 

Photonic lattice is an ideal physical platform to implement continuous-time quantum walk. A large evolution space in the photonic lattice allowing for real spatial two-dimensional quantum walks has been recently demonstrated\cite{Tang2018, Tang2018b}. While this physical system is suitable for coherent and pure quantum walk, the environmental decoherence term can also be intentionally introduced by lattice manipulation. The key process is to introduce classical randomness to the propagation constant along different segments of each waveguide, which causes the randomness in the diagonal part of the Hamiltonian matrix. Therefore, a different kind of evolution, namely, the stochastic quantum walk, has been successfully demonstrated in the photonic lattice to simulate various open quantum systems\cite{Caruso2016, Tang2019}.

\begin{figure}[ht!]
\includegraphics[width=0.46\textwidth]{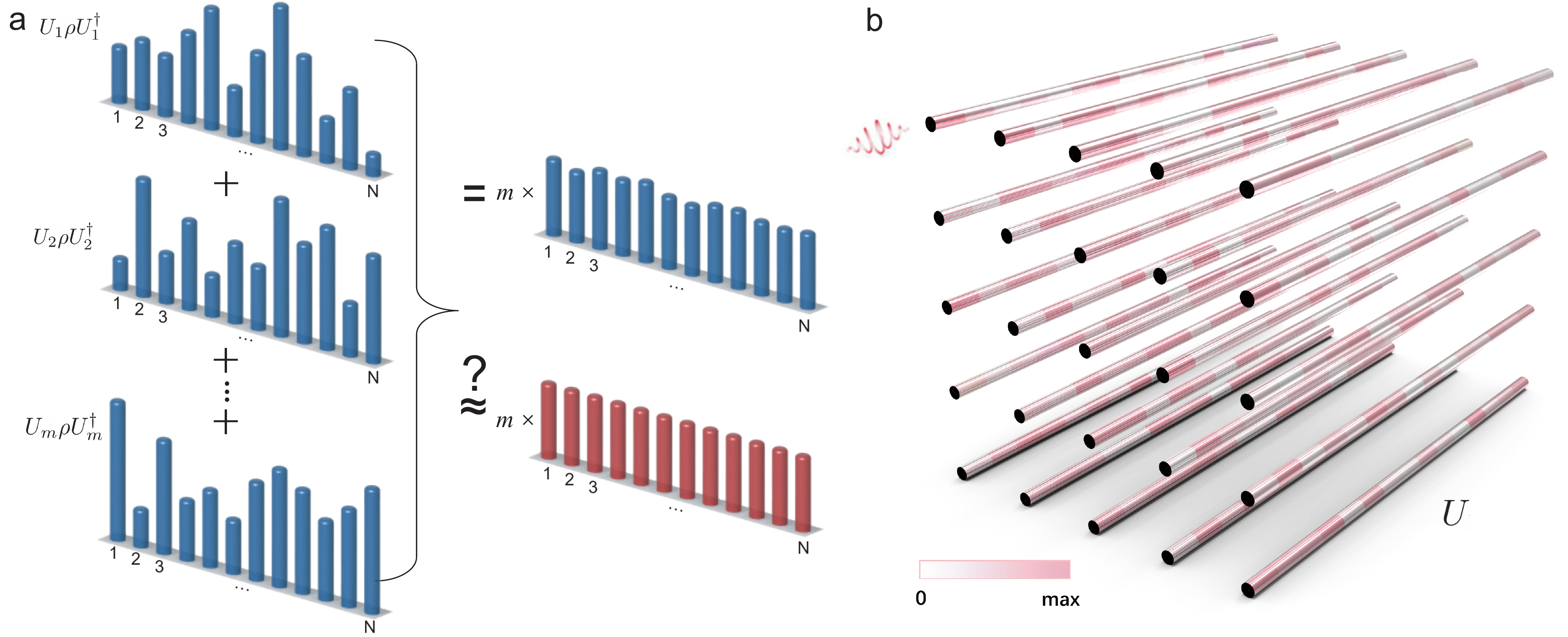}
\caption{\textbf{Generating Haar-uniform randomness using stochastic quantum walks.} ({\bf a}) Illustration of averaging many stochastic quantum walks of a certain evolution time to reach the Haar-measure. The red columns represent the distribution of $I/N$ appeared in Eq.(2). ({\bf b}) Schematic diagram of introducing random delta beta detunings to implement stochastic quantum walks on the photonic chip. The colorbar shows the detuning strength of the propagation constant, where `max' corresponds to the given $\Delta \beta$ amplitude. Photons are injected into one waveguide, and the evolution in the lattice corresponds to a unitary operation. }
\label{fig:Concept}
\end{figure}

In this work, we experimentally demonstrate an instance of Haar-uniform randomness using stochastic quantum walks on the integrated photonic chips. We prepare different samples with different random settings of the propagation constant and detunings, and then measure the light intensity distribution after the evolution inside each chip. The different samples created according to the above procedure yields different unitary evolutions that, in the ideal case, should represent independent samples from the Haar distribution. We show that the average of all distribution profiles converges to the even distribution when the evolution length increases, suggesting the 1-pad Haar-uniform randomness. We further show that our two-dimensional array outperforms the one-dimensional array of the same number of waveguide for the speed of convergence. Additionally, we analytically and numerically show convergence towards uniform distribution when injecting two or multiple indistinguishable photons into such photonic lattice. Our work demonstrates a highly scalable and robust physical implementation for generating Haar-uniform randomness, which can provide useful building blocks to boost future quantum information techniques.

{\it The Experimental Scheme}
We start this section by briefly recalling the most stringent criterion to check when the averages over an ensemble of $N\times N$ unitary operators $\{U_{i}\}$ approximate the formal averages with respect to the Haar distribution. This is normally studied within the framework of approximate $q$-designs\cite{Harrow2009, Banchi2017, Brandao2016} (see Supplementary Note 1 for more details), which require a small distance between two particular averages: 
\begin{equation}
\Vert\mathbb{E}_i[U_{i}^{\bigotimes q}\rho U_{i}^{\bigotimes q\dagger}]-\int_{U} U^{\bigotimes q}\rho U^{\bigotimes q\dagger}dU\Vert_{\diamond}<\varepsilon 
\label{diamain}
\end{equation}
where $\Vert T \Vert_{\diamond}$ is the diamond norm of a superoperator $T$, $\rho$ is the density matrix and $\varepsilon$ is the required small value. $\mathbb{E}_i$ denotes the expectation value, $i.e.$, the average over the ensemble of unitary operators. The second part refers to averages with respect to the Haar measure $dU$. We may also consider a weaker criterion by forcing $\rho$ to be symmetric over permutations, to implement $U_{i}^{\bigotimes q}\rho U_{i}^{\bigotimes q\dagger}$ via experiments involving $q$ indistinguishable photons, where the chip performs the same unitary to each indistinguishable particle.

\begin{figure}[b]
\includegraphics[width=0.45\textwidth]{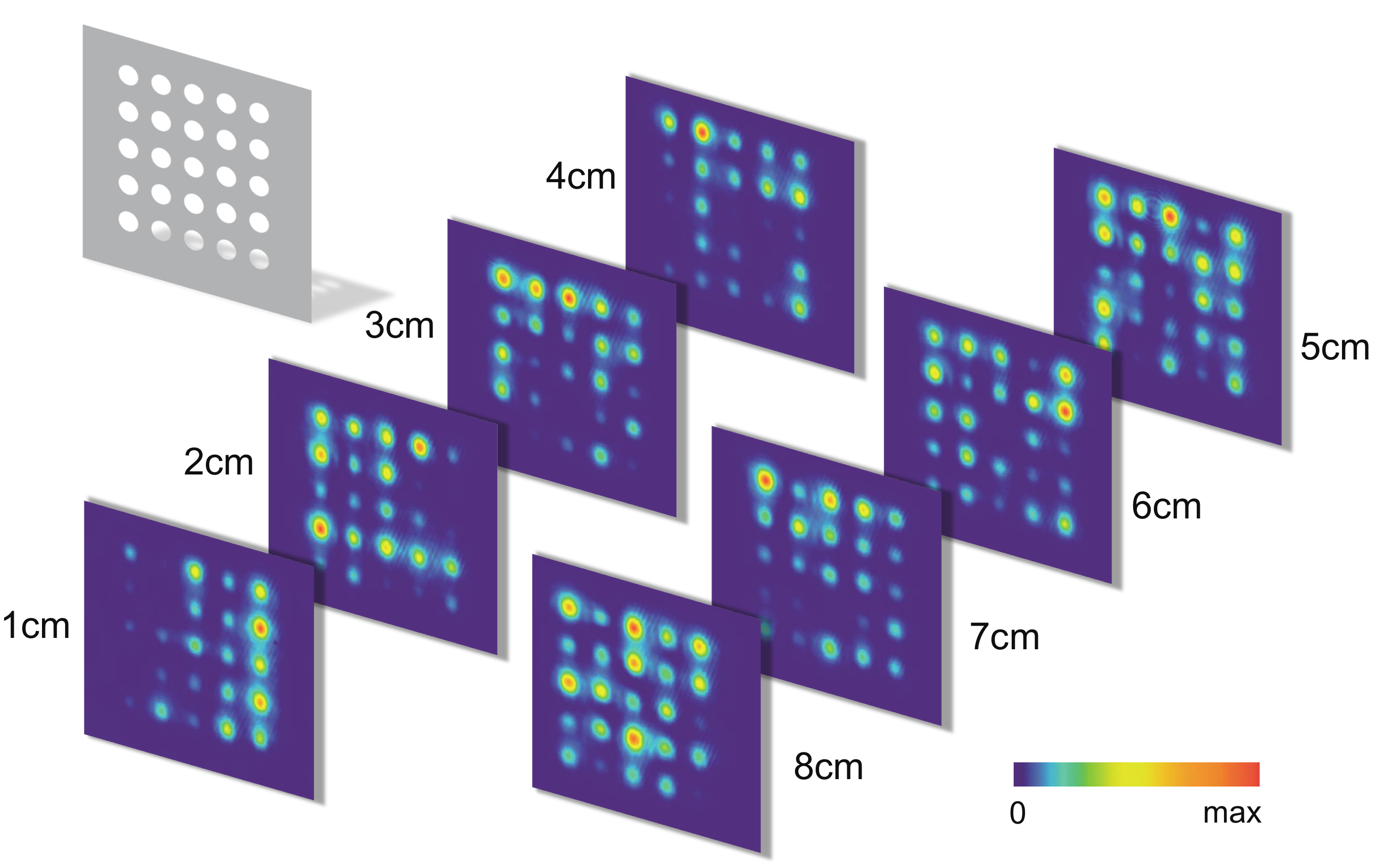}
\caption{\textbf{Experimental results for stochastic quantum walks.} The photonic evolution patterns of different evolution lengths for one random setting of photonic lattice. The corresponding evolution length of each graph is marked beside the graph. The mask illustrates how we read the figure data to get the probability distribution, with details explained in Supplementary Note 5.}
\label{fig:strutturaChip}
\end{figure}

\begin{figure}[t!]
\includegraphics[width=0.5\textwidth]{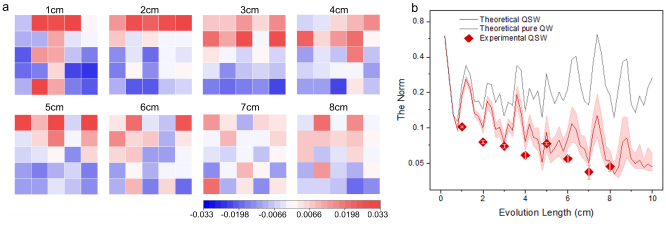}
\caption{\textbf{Convergence to the Haar measure.} (\textbf{a}) The heatmaps show all the elements of the matrix $M$, the diagonal elements of $\mathbb{E}_i[U_{i}\rho U_{i}]-I/N$, for different evolution lengths $z$ as shown above each heatmap. The average for each $z$ is estimated from 17 random settings with the same evolution length. Note that a few elements have a value above 0.033 or below -0.033. They are represented in the heatmap using the color for 0.033 or -0.033, respectively. (\textbf{b}) The L2 norm $\Vert M \Vert$ for samples of different evolution lengths. QSW and QW stand for stochastic quantum walk and quantum walk, respectively. The theoretical results are obtained by setting $\Delta z$ of 2mm and averaging 17 random settings, which are consistent with the experiment. The shading area shows a possible range of theoretical results considering a 15$\%$ fluctuation of the $\Delta \beta$ amplitude, that is, the upper bound and lower bound of the shading area correspond to a $\Delta \beta$ amplitude of 0.34${\rm mm}^{-1}$ and 0.46${\rm mm}^{-1}$, respectively. The error bar for the experimental results is the standard error of the mean, with details explained in Supplementary Note 6.}
\label{fig:apparato}
\end{figure}

Here we firstly focus on $q=1$, which has important applications for quantum encryption\cite{Boykin2003}, and we will briefly analyze $q>1$ scenarios in later sections. The estimation of the diamond norm in Eq.~\eqref{diamain} requires entangled photonic inputs and complete tomography, which is challenging for experiments, as the measurement of each off-diagonal element requires different optical components\cite{Pitsios}. For this reason, in Supplementary Note 1, we introduce a weaker condition based on the L2-norm for which one can prove that entangled inputs are not required. Eq.(S11) links the theoretical norm of $\mathbb{E}_i[U_{i}\rho U_{i}^{\dagger}]-I/N$ to our experimentally measured norm for $M$, namely the distance between the diagonal elements and the uniform distribution (denoted as $\mathcal N_d$). There is another term in Eq.(S11) due to the off-diagonal part $\mathcal N_{od}$. The experiment for measuring the off-diagonal elements of $\mathbb{E}_i[U_{i}\rho U_{i}^{\dagger}]$ would be too complex, and we manage to show in Supplementary Note 2 and Fig.S1 numerically that the off-diagonal elements for the photonic lattice model used in this work asymptotically converge to zero. Hence we focus on the experimental exploration on diagonal elements which directly corresponds to the probability distribution at each waveguide, and would experimentally verify whether these diagonal terms converge to zero. 

For $q=1$, the Haar average reduces to $I/N$, where $N$ is the number of waveguides and $I$ is the $N\times N$ identity matrix. Therefore, Eq.(1) demands:
\begin{equation}
\Vert\mathbb{E}_i[U_{i}\rho U_{i}^{\dagger}]-I/N\Vert_d<\varepsilon 
\end{equation}
for all input states $\rho$, where $\|\cdot\|_d$ is the L2 norm of the diagonal elements. This is inspiring from the experimental perspective, as illustrated in Fig.1a. However, not all unitary ensembles can satisfy Eq.(2). For instance, quantum walks with a pure state input and an ensemble made by a single unitary have a fixed ballistic distribution rather than a uniform one. On the other hand, the theoretical proposal\cite{Banchi2017} shows that averages over different continous-time stochastic quantum walks with sufficiently long waveguides could successfully approximate the Haar averages. 

Consider a photonic lattice where each waveguide is equally divided into the same number of segments, and each segment has a constant detuning of the propagation constant, with random detunings in all segments following a uniform distribution (See Fig.1b). The photon evolution through such a lattice with $N$ waveguides corresponds to the operation with $U$ of a size $N$. The evolution can be described by an effective piecewise Hamiltonian $H_{\rm eff}$. For each segment $k$, there is:
\begin{equation}      
H_{\rm eff}(k)= \sum_{i}^N (\beta_i+\Delta \beta_{i}(k))a_{i}^\dagger a_{i}+\sum_{j\neq i}^N C_{ij}(a_{i}^\dagger a_{j}+a_{j}^\dagger a_{i})
\end{equation}
where $\beta_i$ and $C_{ij}$ are respectively the propagation constant and coupling coefficient for the lattice without any detunings. In practice, $\beta_i$ is set the same for all waveguides. $\Delta \beta_{i}(k)$ is the constant detuning of the propagation constant for waveguide $i$ at segment $k$, which can be experimentally achieved by tuning the writing speed (See details for $\Delta \beta$ tuning in Supplementary Note 3 and Fig.S2). The introduction of $\Delta \beta$ in $H_{\rm eff}$ has an effect of adding some classical dephasing terms during the evolution.

Thanks to Eq.(3), the unitary implemented by a chip of length $z=K\Delta z$ is $U(z)=(\prod_k e^{-iH_{\rm eff}(k)\Delta z})$, where $\Delta z$ is the length of each segment, and $K$ is the number of segments. Calling $\ket{\Psi(z)}=U(z)\ket{\Psi(0)}$ for a given initial wavefunction $\ket{\Psi(0)}$, what is experimentally measured is the probability distribution $|\braket {l}{\Psi(z)}|^2$ that the photon will come out from the $l$th waveguide after an evolution length $z$. The probability distribution is equivalent to the diagonal elements of $U\rho U^{\dagger}$.

In the experiment, we prepare photonic lattices of 5$\times$5 waveguides, a total evolution length of 8cm, and a segment length $\Delta z$ of 2mm. The random $\Delta \beta$ detunings in all segments follow a uniform distribution under a $\Delta \beta$ amplitude of 0.4${\rm mm}^{-1}$ using femtosecond laser direct writing tecchnique\cite{Tang2018,Crespi2013,Chaboyer2015}(See details for waveguide preparation in Supplementary Note 4). In total, we have 17 random settings for the detuning profiles. We inject photons from one waveguide of the lattice and measure the evolution patterns for an evolution length of 1cm, 2cm, 3cm, 4cm, 5cm, 6cm, 7cm, and 8cm, which will allow us to see how the performance changes with the evolution length. The experimental evolution patterns are given in Supplementary Fig.S3 - S10.

\begin{figure}[t!]
\includegraphics[width=0.4\textwidth]{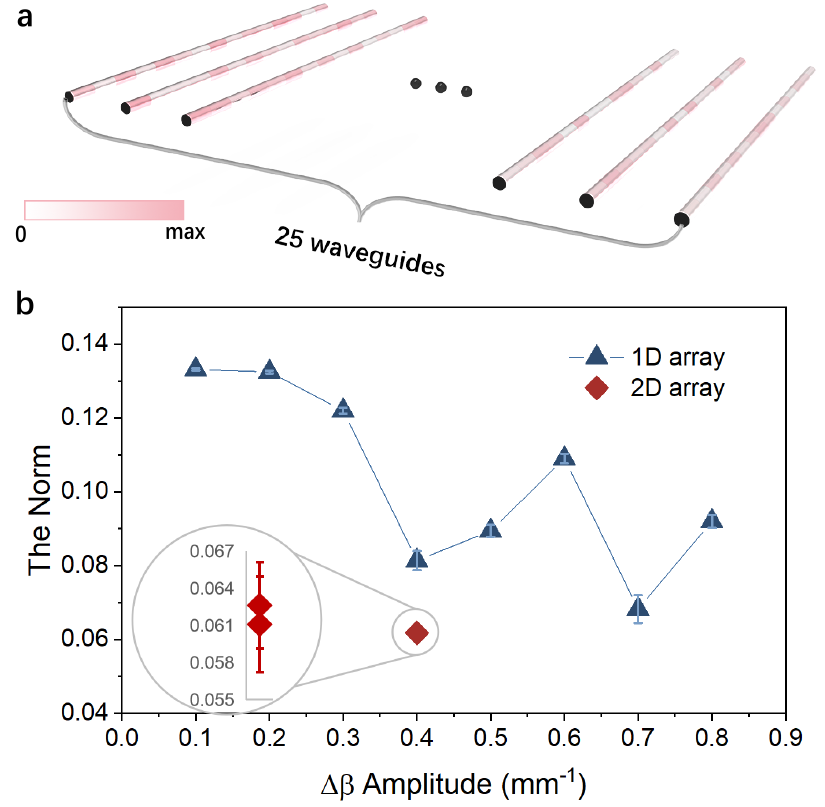}
\caption{\textbf{Compare the performance in one- and two-dimensional array.} (\textbf{a}) Schematic diagram of the one-dimensional photonic lattice of 25 waveguides with random tunings of the propagation constant. (\textbf{b}) The calculated L2 norm $\Vert M \Vert$ for one-dimensional (1D) and two-dimensional (2D) arrays. For two-dimensional array, we get two sets of norm values, each by averaging 6 random settings separately.   }
\label{fig:Results4}
\end{figure}

{\it Result Analysis} As shown in Fig.2, we measure the photonic evolution pattern for different evolution lengths after injecting a photon in the lattice of one random setting. We then read the intensity probability at each waveguide for each figure (See details in Supplementary Note 5). We have processed all 17 random settings and each has 8 different evolution lengths.  

For each evolution length, we average the probability distribution of the 17 settings. What we obtain is the diagonal part of $\mathbb{E}_i[U_{i}\rho U_{i}^{\dagger}]$ in Eq.(2). The diagonal part of the other term in Eq.(2), $I/N$, can be viewed as the equal distribution at all 5$\times$5 waveguides, which means each waveguide has an equal probability of 0.04. We substract 0.04 from each element of the measured average probability distribution matrix, and we can get the diagonal part of $\mathbb{E}_i[U_{i}\rho U_{i}^{\dagger}]-I/N$, which is a 25$\times$ 1 vector and can be written in a $5\times5$ matrix $M$. 

We use the heatmap to list the value of each element in the matrix $M$ (See Fig.3a). Clearly, for a small evolution length, the fluctation around zero for these element values is much more fierce than that for a larger evolution length. We calculate the L2 norm of $M$, $i.e.$, $\Vert M \Vert$. If all elements in $M$ are zero, $\Vert M \Vert$ will certainly be zero, while large deviations from zero in these elements make $\Vert M \Vert$ big. The calculated $\Vert M \Vert$ shown in Fig.3b well supports the results in heatmaps. $\Vert M \Vert$ gradually converges to a considerably small value when the evolution length increases. A slight gap between the experimental and numerical results may due to imperfect fitting for $\Delta \beta$ as explained in Supplementary Note 3, but overall there is a good match to show that stochastic quantum walk results dynamically decay to zero.  As a comparison, the pure quantum walk always keeps a high norm value and does not show a sign of convergence. 

Our experiment demonstrates that the diagonal elements of $\mathbb{E}_i[U_{i}\rho U_{i}^{\dagger}]$ indeed tend towards a uniform distribution. We further show in Supplementary Note 7 and Fig.S11-S12 that, within a certain evolution length scale, increasing the evolution length lowers down the norm, before approaching a constant trend. Increasing the $\Delta \beta$ amplitude can speed up the convergence without altering the norm lower bound, while increasing the number of samples can further lower down the norm. In Fig.S13 we show the segment length $\Delta z$ also influences the convergence length without changing the norm lower bound.   The convergence length does not monotonically change with $\Delta z$, but reaches a minimal value when choosing a proper $\Delta z$ value. Our experiment demonstration of the diagonal elements, together with the numerical convergence of off-diagonal part in Supplementary Note 2, show that stochastic quantum walks can reach a 1-design Haar measure at a long enough evolution length. 

We further investigate the one-dimensional photonic lattice of 1$\times$25 waveguides (see Fig.4a), a segment length $\Delta z$ of 2mm and an evolution length of 8cm. We set 8 different $\Delta \beta$ amplitudes, namely, 0.1, 0.2, 0.3, 0.4, 0.5, 0.6, 0.7 and 0.8 ${\rm mm}^{-1}$, and each has 6 random samples. The experimental patterns for all one-dimensional samples are provided in Supplementary Fig.S14. We average the 6 probability distributions for each $\Delta \beta$ amplitude and plot their $\Vert M \Vert$ in Fig.4b. For the one-dimensional array, as $\Delta \beta$ amplitude increases, there is a slightly reducing trend of the norm. We show in Fig.S15 a similar trend for the two-dimensional lattices, that they converge better at a higher $\Delta \beta$ amplitude. This is because a stronger dephasing effect caused by larger $\Delta \beta$ detunings can facilitate a faster convergence to the Haar measure. However, the norm for one-dimensional samples of large $\Delta \beta$ amplitudes still exceeds the norm for the two-dimensional samples with a $\Delta \beta$ amplitude of 0.4${\rm mm}^{-1}$. The two-dimensional quantum walk has demonstrated the same ballistic transport with the one-dimensional quantum walk, and yet a faster decay from the injection site than the latter. This is due to richer evolution paths\cite{Tang2018} that also allow for more flexible Hamiltonian engineering\cite{Tang2018b}. In this work, we show that the two-dimensional stochastic quantum walk has a clear advantage in fast convergence to the Haar measure utilizing the rich evolution paths. 

In addition, we explore the convergence to Haar measure for the scenario of $q>1$ via some analytical and numerical analysis, as shown in Supplementary Note 8 and Fig.S15. The probability distribution when injecting two indistinguishable photons from mode $i$ and $i'$, would converge to $\frac2{N(N+1)}$. Alternatively, the compound light intensity at each mode $j$, $I_{j}$, would converge to $\frac2{N}$. The convergence follows the same dependence on the $\Delta \beta$ amplitude and on the number of groups as in $q=1$. Meanwhile, both $q=2$ and $q=1$ converge at a similar evolution length, showing the same convergence speed, as theoretically expected\cite{Banchi2017}. The difference lies in that the lower bound for $q=2$ is slightly higher. We further derive a general expression for the average probability, that is, $\prod_{k=1}^{q} \frac{k}{N+q-k}$, and $\frac{q}{N}$ for the compound light intensity. They can be reduced to the aforementioned expression for $q=1$ or 2.

{\it Discussion}
By taking full advantages of large-scale integrated photonic chips and precise lattice manipulation techniques, we manage to use the classical randomness ($i.e.$, the random settings of $\Delta \beta$ detunings) to generate an important source for quantum randomness and make the unitary 1-design distribution. That is, we have been able to demonstrate large-scale continuous-time stochastic quantum walks on the photonic chips, and suggest the convergence to Haar-uniform randomness at a large evolution length. 

Our approach offers a highly feasible alternative to the quantum gate approach or the Reck/Clements decomposition approach for generating Haar randomness. The Reck/Clements scheme offers a way of decomposing any certain unitaries to avoid the difficulties of constructing different configurations\cite{Harris2017}, and yet it requires fine tuning of a quadratic number of parameters of beamsplitters and phase shifters as the unitary scales up.

On the other hand, our approach does not need fine tuning or complicated quantum circuit designs. Indeed, as long as the quantum walk Hamiltonian is ``fully-controllable'', namely any unitary is reachable with proper local phases, random $\Delta\beta$s with suitably long chips eventually yield Haar random unitary evolutions, without the need of precise phase calibration. Slightly different $\Delta \beta$ settings are mapped in fluctuations in the resulting unitary, which would also follow a Haar measure. Experimentally, we programmed the random $\Delta\beta$ with femtosecond laser writing speed. Even if the scale of detunings (the $\Delta \beta$ amplitude) is not kept as desired, after a long enough evolution (shown in Fig.S12), the average will still converge to the Haar measure. Meanwhile, the convergence can be improved with larger $\Delta \beta$ detunings, a properly selected segment length $\Delta z$, and more samples of different $\Delta \beta$ detunings. From Fig.S12-13, we see that, a $\Delta\beta$ amplitude of $0.6 {\rm mm}^{-1}$ with a $\Delta z$ of 2mm, or a $\Delta\beta$ amplitude of $0.4 {\rm mm}^{-1}$ with a $\Delta z$ of 5mm, can both converge at a feasible evolution length below 10cm. Therefore, we need to choose proper and feasible parameters considering today's fabrication technologies. Being free of fine tuning and having feasible parameters are advantageous features for our approach.

In this work, we used many samples to compare ensemble averages with theoretical averages, and proved convergence towards the Haar distribution. Nonetheless, a single Haar-random chip is enough for most applications, e.g. boson sampling: once it is understood that noisy quantum walks converge towards the Haar distribution, in applications it is enough to fabricate a single chip, designed according to the theoretical recipe. Our experiment focuses on proving that noisy quantum walks do indeed converge towards the Haar distribution, and additionally our analytical results show the convergence does apply to scenarios of multiple photons. Our utilization of two-dimensional continuous-time stochastic quantum walk on photonic chips sheds light for more applications that need Haar randomness.

\begin{acknowledgements}
{\it Acknowledgments}
The authors thank Jian-Wei Pan for helpful discussions. We thank Zi-Yu Shi for her previous characterization on $\Delta\beta$ together with H.T., which is helpful for this work. This research was supported by National Key R\&D Program of China (2019YFA0706302,2019YFA0308700, 2017YFA0303700), National Natural Science Foundation of China (61734005, 11761141014, 11690033, 11904229), Science and Technology Commission of Shanghai Municipality (STCSM) (17JC1400403), and Shanghai Municipal Education Commission (SMEC) (2017-01-07-00-02- E00049). X.-M.J. acknowledges additional support from a Shanghai talent program. MSK's work is supported by the UK Hub in Quantum Computing and Simulation with funding from UKRI EPSRC grant EP/T001062/1 and a Samsung GRP grant. L.B. acknowledges support by the program ``Rita Levi Montalcini'' for young researchers.
\end{acknowledgements}

\bigskip

\setcounter{equation}{0}
\setcounter{figure}{0}
\renewcommand{\theequation}{{S}\arabic{equation}}
\renewcommand{\thefigure}{{S}\arabic{figure}}

\subsection{Supplementary Note 1: Remarks on the convergence towards the completely mixing evolution}

As discussed in the main text, for $q=1$ we may formally study convergence 
towards a fully mixing evolution via the 
diamond distance $\|\mathcal E_{\rm ensemble}-\mathcal E_{\rm Haar}\|_\diamond$
between the quantum channel that mathematically describes
the average over $m$ discrete unitaries, $\mathcal E_{\rm ensemble}[\rho] =\sum_{j=1}^m U_j \rho U_j^\dagger$, 
and the quantum channel that describes continuous averages over Haar-random unitaries 
$\mathcal E_{\rm Haar}[\rho] =\int dU \, U \rho U^\dagger$. Such distance can be written as 
(see {\it e.g.}~\cite{Watrous2018}) 
\begin{equation}
	\|\mathcal E_{\rm ensemble} -\mathcal E_{\rm Haar}\|_\diamond  = 
	\max_\Psi \| (\openone\otimes\mathcal E_{\rm ensemble}-\openone\otimes\mathcal E_{\rm Haar})(\Psi)\|_1
	\label{diamond},
\end{equation}
where the maximization is over quantum states $\Psi=\ket\Psi\!\bra\Psi$. Although such maximization can be efficiently solved 
using semidefinite programming for small dimensional Hilbert spaces, it is generally complex to estimate the 
diamond distance from measurement data. Therefore, we study simpler quantities that are related to the diamond distance via bounds.

We first recall one of the possible definitions of the trace norm (see {\it e.g.}~\cite{Bengtsson2017}
Theorem 13.2)
\begin{equation}
	\|\rho-\sigma\|_1 = \max_{E} \sum_i |\Tr[E_i (\rho-\sigma)]|,
	\label{helstrom}
\end{equation}
where the maximum is taken over all possible POVMs such that $\sum_i E_i = \openone$. 
With the above definition, we may introduce the following inequality
\begin{equation}
	\|\mathcal E_{\rm ensemble} -\mathcal E_{\rm Haar}\|_\diamond 
	\le N \| \chi_{{\rm ensemble}}-\chi_{{\rm Haar}}\|_1~,
	\label{diamondineq}
\end{equation}
where $\chi_{t}=\openone\otimes\mathcal E_t[\ket\Phi\!\bra\Phi]$ is the Choi-Jamio{\l}kowski state 
associated to the channel $\mathcal E_t$ and $\ket\Phi=\sum_{j=1}^N\ket{jj}/N$.
Indeed, using \eqref{helstrom} in \eqref{diamond} we may write for two generic channels,
$\mathcal E_0$ and $\mathcal E_1$, the following inequality
(also discussed in \cite{Watrous2018} Exercise 3.6)
\begin{align*}
	\|\mathcal E_0 -\mathcal E_1\|_\diamond  &= 
	\max_\Psi \| (\openone\otimes\mathcal E_0-\openone\otimes\mathcal E_1)(\Psi)\|_1
	\cr &=
	\max_{E,\Psi} \sum_i |\Tr \left[E_i(\openone\otimes\mathcal E_0-\openone\otimes\mathcal E_1)(\Psi)\right]|
	\cr &=
	\max_{E,M} \sum_i |\Tr \left[E_i(M\otimes \openone)(\chi_{0}-\chi_{1})M^\dagger\otimes\openone\right]|
	\cr &=
	\max_{M} \|(M\otimes \openone)(\chi_{0}-\chi_{1})M^\dagger\otimes\openone\|_1
	\cr &\le \max_M \|M\|_\infty^2 \|\chi_{0}-\chi_{1}\|_1 
	\cr &\le \max_M \|M\|_2^2 \|\chi_{0}-\chi_{1}\|_1 
	\cr &= N \|\chi_{0}-\chi_{1}\|_1 ,
\end{align*}
where, without loss of generality, we have set $\ket{\Psi} ={M\otimes\openone}\ket\Phi$, so that
$M$ satisfies $\|M\|_2^2=\Tr M^\dagger M=N$ due to normalization. Indeed, from $\ket{\Psi} = 
\sum_{ij} \Psi_{ij} \ket{ij}$ we may set $M_{ij}=\sqrt N \Psi_{ij}$. In the fifth line 
we use the property $\|ABC\|_1 \le \|A\|_\infty \|B\|_1\|C\|_\infty$ (see \cite{Watrous2018}
section 2.3.1), and in sixth line $\|A\|_\infty \le \|A\|_2$. 

The trace distance in Eq.~\eqref{diamondineq} could be measured experimentally as follows. 
The state $\chi_{{\rm ensemble}}$ may be created by first generating path-entangled 
photonic states, and then sending half of the entangled state inside the linear-optical network 
that implements a quantum walk (or any other unitary). On the other hand, using simple 
properties of Haar integrals we find $\mathcal \chi_{{\rm Haar}}=\openone/N^2$. 
The resulting  experiment, although technically possible,
is made more challenging by the use of entangled photon sources and by the demand of complete 
tomography, for reconstructing $\chi_{{\rm ensemble}}$.
In order to introduce a simpler quantity for experiments we expand $\chi_{\rm ensemble}$ 
into the basis $\ket i$, where $\ket i$ means that a photon is injected (or detected) in 
the $i$th waveguide. Using the definition of the Choi-Jamio{\l}kowski state, we may write 
\begin{align}
	\chi_{\rm ensemble} &= \sum_{ij,kl} \frac1N \bra k \mathcal E_{\rm ensemble}[\ket i\!\bra j]\ket l ~ \ket{ik}\!\bra{jl} = 
	\nonumber
	\\ &= \sum_{ij,kl} \chi_{ik,jl} \ket{ik}\!\bra{jl}~,
\end{align}
where we have defined  $ \chi_{ik,jl} = \bra k 
\mathcal E_{\rm ensemble}[\ket i\!\bra j]\ket l/N$. We may now split 
\begin{equation}
	\chi_{\rm ensemble} = \chi_{\rm d} + \chi_{\rm od},
\end{equation}
where $\chi_{\rm d} = \sum_{ik} \chi_{ik,ik} \ket{ik}\!\bra{ik}$ is the diagonal part or $\chi_{\rm ensemble}$
and $\chi_{\rm od}$ its off-diagonal part. 
Therefore, using the bound \eqref{diamondineq} and the triangle inequality we may write 
\begin{equation}
	\|\mathcal E_{\rm ensemble} -\mathcal E_{\rm Haar}\|_\diamond 
	\le N (D_{\rm d} + D_{\rm od})~,
\end{equation}
where $D_{\rm d}$ measures the distance of the diagonal elements from the uniform distribution,
\begin{align}
	D_{\rm d} &= \|\chi_d-\openone/N^2\|_1 = \sum_{ik} \left|\chi_{ik,ik}-\frac1{N^2}\right| = 
					\\&= \sum_{ik} \frac1N \left|\sum_{l=1}^m \bra k U_l \ket i\!\bra i U_l^\dagger \ket k -\frac1N\right| =
					\\&= \frac1N \sum_{i,k=1}^N \left|\sum_{l=1}^m |\bra k U_l \ket i|^2  -\frac1N\right|,
\end{align}
and $D_{\rm od}=\|\chi_{\rm od}\|_1 \leq 2$ measures the strengths of the off-diagonal elements. 

In order to define a simpler quantity for experimental tests, we consider focus on 
\begin{equation}
	\mathcal N = 
	\max_\Psi \| (\openone\otimes\mathcal E_{\rm ensemble}-\openone\otimes\mathcal E_{\rm Haar})(\Psi)\|_2,
	\label{diamond2}
\end{equation}
which is analogous to Eq.~\eqref{diamond}, but with the $\ell_2$-norm in place of the $\ell_1$-norm. 
The maximization in the above norm is over all possible input states but, due to convexity of matrix norms, 
we may restrict such maximization to pure states. Indeed, for any convex function $f(\rho)$ of a quantum state $\rho$, 
we may write $f(\sum_i p_i \ket{i}\!\bra{i})\leq \sum_i p_i f(\ket{i}\! \bra{i})\leq  \max_i f(\ket{i}\!\bra{i})$,
so the maximum is always obtained with pure states. Without loss of generality, we can also choose 
$\ket{\Psi} = \sqrt{\rho}\otimes\openone\ket{\Phi}$, namely as a purification of the state $\rho$. 
Then it is simple to show that $\mathcal N^2 = \Tr[(\rho\otimes\openone)\chi(\rho\otimes\openone)\chi]$, 
where $\chi=\chi_{\rm ensemble}-\chi_{\rm Haar}$, which 
is convex over $\rho$. As such the maximum is obtained for pure $\rho$ or, in other terms, when 
$\Psi$ in Eq.~\eqref{diamond2} is separable. Since $\mathcal E_{\rm Haar}(\Psi)=\openone/N$ we can write
\begin{equation}
	\mathcal N^2 = \max_{\ket{\psi}} \Tr[\mathcal E^2_{\rm ensemble}(\ket\psi\!\bra\psi)]-\frac1N
	= \max_{\ket\psi}\left[\mathcal N_{\rm d}^2 + \mathcal N^2_{\rm od}\right],
\end{equation}
where 
\begin{align}
	\mathcal	N_{\rm d}^2 &= \sum_k \left[p_{\ket\psi}(k)-\frac1N\right]^2~,\\
	\mathcal	N_{\rm od}^2 &= \sum_{k,l:k\neq l}  |\bra k\mathcal E_{\rm ensemble}(\ket\psi\!\bra\psi)\ket l|^2,
\end{align}
are respectively the diagonal and off-diagonal contribution and 
$p_{\ket\psi}(k) = |\bra k\mathcal E_{\rm ensemble}(\ket\psi\!\bra\psi)\ket k|$. 

\subsection{Supplementary Note 2: The numerical analysis of the off-diagonal elements}

We consider the $\Delta \beta$ approach that produces the effective Hamiltonian $H_{\rm eff}$ given in Eq.(3) of the main text. The evolution of such stochastic quantum walks implement the operation $U\rho U^{\dagger}$. Experimentally, the diagaonal elements of $U\rho U^{\dagger}$ can be measured by the probability distribution, while the off-diagonal elements are difficult to measure, but still can be numerically calculated using the $\Delta \beta$ model. 

We consider the parameter setting of $\beta_0$ and $C_{ij}$ to be consistent with those in the experiment. The square waveguide lattice has a nearest-neighbor coupling coefficient of 0.225 $\rm mm^{-1}$, and a next-nearest-neighbor coupling coefficient of around 0.09 $\rm mm^{-1}$, as has been characterized for the same waveguide spacing configuration in \cite{Tang2018}. For waveguides further apart, the coupling coefficients are small enough to be not considered. The absolute value of $\beta_0$ does not influence the numerical result. In this work, we use a $\beta_0$ value of 11660 $\rm mm^{-1}$.

Here we calculate the norm for the off-diagonal elements of $\mathbb{E}_i[U_{i}\rho U_{i}^{\dagger}]$ for different $\Delta \beta$ amplitudes, the scale of dephasing, and different numbers of samples. As shown in Fig. S1, increasing the number of samples and the $\Delta \beta$ amplitude could clearly improve the convergence to zero. This suggests that the off-diagonal part of the stochastic quantum walk in the photonic lattice model proposed in this work can reach the uniform distribution to meet the Haar randomness measure. 

\subsection{Supplementary Note 3: The $\Delta \beta$ photonic approach} 
Before introducing the method of $\Delta \beta$ generation, we would firstly explain how to generally measure the coupling coefficient $C$ using the directional coupler approach~\cite{Szameit2007}. In the directional coupler, light intensity in the two waveguides are denoted as $I(A)$ and $I(B)$, respectively. The intensity would vary sinusoidally with the evolution length $z$, that is, $I(A)=|{\rm cos}(Cz)|^2$, $I(B)=|i {\rm sin}(Cz)|^2$. Therefore,
\begin{equation}
	\frac{I(A)-I(B)}{I(A)+I(B)}=\frac{{\rm cos}^2(Cz)-{\rm sin}^2(Cz)}{{\rm cos}^2(Cz)+{\rm sin}^2(Cz)}={\rm cos}(2Cz).
\end{equation}
Experimentally, we prepare a group of directional couplers of different evolution lengths, and measure the light intensity at each of the two modes (see Fig.S2a). Note that we just use two straight waveguide to form the directional couplers and it does not have a curved part. This avoids unnecessary loss in the bending part and avoids the unwanted coupling in the bending region where two waveguides gradually go apart. We can fit the relationship between $\frac{I(A)-I(B)}{I(A)+I(B)}$ and $z$ with a suitable sinusoidal curve (see Fig.S2b), and the coupling coefficient $C$ can be obtained from the frequency of this sinusoidal curve.

Then $\Delta \beta$ can be measured using this directional coupler approach. One of the waveguide is written using a base speed $V_0=5~\rm mm/s$, and the other waveguide using a different detuned speed $V$ ($V-V_0=\Delta V$) that will lead to a detuned propagation constant $\Delta \beta$ on this waveguide. In the detuned directional coupler, the effective coupling coefficient $C_{\rm eff}$ can be obtained using the same coupling mode method as that for the normal directional coupler~\cite{Szameit2007}, but $C_{\rm eff}$ contains the detuning effect from $\Delta \beta$ through this equation~\cite{Lebugle2015}: 
\begin{equation}
C_{\rm eff}=\sqrt{(\Delta \beta/2)^2+C_0^2},
\end{equation}
where $C_0$ is the coupling coefficient for a normal directional coupler. Therefore, $\Delta \beta$ can be calculated when $C_{\rm eff}$ and $C_0$ are both characterized. 

We prepared 13 groups of samples with a base speed of $5~\rm mm/s$ and a detuned speed ranging from 5, 7.5, 10, 12.5, all the way to 35$~\rm mm/s$. This corresponds to a $\Delta V$ from 0 to $30~\rm mm/s$. Each group has 12 samples of the same speed setting but different evolution lengths such that we can fit the sinusoidal curve to get $C_{\rm eff}$ or $C_0$ (when $\Delta V=0$).
With these 156 samples, we then obtain $\Delta \beta$ (unit: $\rm mm^{-1}$) against $\Delta V$ (unit: $\rm mm/s$), as ploted in Fig.S2c. There can be a rough fit linearly: $\Delta \beta=0.02\times \Delta V$. The linear relationship is consistent with a previous report\cite{Caruso2016}. Knowing this, we can randomly generate $\Delta \beta$ of $0.01-0.4~\rm mm^{-1}$ by varying $\Delta V$ between $0.5-20~\rm mm/s$. 

It is worth mentioning that in this experiment, we set a target $\Delta \beta$ value by strictly following the linear relationship with $\Delta V$. This might cause some systematic error as the real generated $\Delta \beta$ might go above  $0.02\times \Delta V$, e.g., a $\Delta V$ of $20~\rm mm/s$ might have generated a $\Delta \beta$ of $0.5~\rm mm^{-1}$ instead of $0.4~\rm mm^{-1}$. This may explain the slightly better convergence in experiment than in numerical analysis in Fig. 3b of the main text. Nonetheless, there is a good match of the convergence trend overall. 

It is also worth noting that although we calibrate $\Delta \beta$, an absolute value of $\beta$ detuning, an increase in writing speed in fact corresponds to a decrease in $\beta$. In this work, plus or minus a $\beta$ detuning has almost no influence in the numerical analysis, but the sign may matter for many other work that require $\beta$ detunings such as the manipulation of site energies, so we believe it's necessary to clarify this for wide readers. 

Additionally, our experimental limit for generating a $\Delta \beta$ can reach around $1.0~\rm mm^{-1}$ that requires a speed detuning of around $50~\rm mm/s$. This allows for rich manipulation of $\Delta \beta$ and can make a convergence at a feasible length on current photoic chips. 

\subsection{Supplementary Note 4: Waveguide preparation}
For the 17 random settings, each has 8 different evolution lengths. As it is not convenient to measure the evolution patterns in the middle of the waveguide, we have to make 8 samples with a length of 1cm, 2cm, ..., 8cm, respectively. We have ensured the 8 samples follow the same random setting, for instance, $\Delta \beta$ profiles for the 4-cm-long sample are exactly the same with those in the first 4cm of the 8-cm-long sample.

Each sample, regardless of the difference in evolution length or random setting, has the same configuration, that is, a 5 $\times$5 square lattice with the nearest waveguide spacing of 15 $\rm \mu m$ in the vertical direction, and of 13.5 $\rm \mu m$ in the horizontal direction. This ensures the nearest-neighbor coupling coefficient to be consistent in different directions\cite{Tang2018}.

All the waveguides are fabricated using the femtosecond laser direct writing technique\cite{Crespi2013, Chaboyer2015, Tang2018}. We direct a 513-nm femtosecond laser (up converted from a pump laser of 10W, 1026nm, 290fs pulse duration, 1 MHz repetition rate) into a spatial light modulator (SLM) to shape the laser pulse in the temporal and spatial domain. The writing speed is precisely controlled for each segment to ensure the introduction of the target $\Delta \beta$ value. We then focus the pulse onto a pure borosilicate substrate with a 50X objective lens (numerical aperture:~0.55). Power and SLM compensation were processed to ensure the waveguide uniformity. 

\subsection{Supplementary Note 5: Measure probability distribution from graphs.} All the experimentally measured evolution patterns with 17 different random settings are provided in Fig.S3-S10. When collecting the data from experiments, we obtained the corresponding ASCII file, which is essentially a matrix of pixels. We created a `mask' containing the pixel coordinate of the circle centre and the radius in pixels for each waveguide, and summed the light intensity for all the pixels within each circle using Matlab. The normalized proportion of light intensity for each circle represents the probability at the corresponding waveguide.

\subsection{Supplementary Note 6: The standard error of the norm}
According to the definition of the norm and our experimental setting, we know how we calculte our norm $N$:
\begin{equation}
	N =\sqrt{\sum_{j=1}^{25}(\frac{\sum_{i=1}^{17}p_{ij}}{17}-0.04)^2},
\end{equation}
where $p_{ij}$ is the probability of $j$th site in $i$th experiment. Replacing it with the mean probability $P_{j}=\frac{1}{17}\sum_{i=1}^{17}p_{ij}$, we have:
\begin{equation}
	N =\sqrt{\sum_{j=1}^{25}(P_{j}-0.04)^2}.
\end{equation}
Using the error propagation formula, we have($\Delta$ means standard deviation):
\begin{equation}
	\Delta N =\sum_{j=1}^{25}(\frac{\partial N}{\partial P_{j}}\Delta P_{j})^2,
\end{equation}
in which $\Delta P_{j}$ is easy to get. Meanwhile: 
\begin{equation}
	\frac{\partial N}{\partial P_{j}}=\frac{P_{j}-0.04}{\sqrt{\sum_{j=1}^{25}(P_{j}-0.04)^2}}=\frac{P_{j}-0.04}{N},
\end{equation}
so totally we have:
\begin{equation}
	\Delta N =\sum_{j=1}^{25}(\frac{P_{j}-0.04}{N}\Delta P_{j})^2.
\end{equation}
Therefore, we can calculate $\Delta N$ knowing $N$, $P_{j}$ and $\Delta P_{j}$. Then the standard error of the mean is $\frac{\Delta N}{\sqrt{17}}$.

As for the theoretical result of 17 random samples, we can also similarly work out a standard error, but we consider there is a more influential factor, that is, we may not exactly fabricate a sample with a $\Delta \beta$ amplitude of 0.4${\rm mm}^{-1}$. Therefore, we set a possible range of theoretical results considering a 15$\%$ fluctuation of the $\Delta \beta$ amplitude. We show a shading area with its upper bound and lower bound corresponding to a $\Delta \beta$ amplitude of 0.34${\rm mm}^{-1}$ and 0.46${\rm mm}^{-1}$, respectively. When calculating the shading area, the random setting configuration is kept the same with that for $\Delta \beta$ amplitude of 0.4${\rm mm}^{-1}$, but just proportionally shrink it to 0.34${\rm mm}^{-1}$ or enlarge it to 0.46${\rm mm}^{-1}$, respectively. 

\subsection{Supplementary Note 7: The norm of $M$ using $\Delta \beta$ photonic approach for a long evolution length}
As shown in Fig.S11, for a longer evolution length, with a $\Delta \beta$ amplitude of 0.4 $mm^{-1}$, the norm of $M$ for stochastic quantum walk becomes closer to zero. We further show in Fig.S12 that increasing the evolution length can effectively reduce the norm to an extremely small value and the norm would get stable after a certain evolution length. We see in Fig.S12 among lattices of different $\Delta \beta$ amplitudes, the one with a higher $\Delta \beta$ amplitude has faster convergence to the Haar measure, but given the same number of sample random settings, lattices show a consistent threshold of the norm regardless of the $\Delta \beta$ amplitude values. Comparing Fig.S4a and 4b that use 100 and 500 random settings, respectively, we see that increasing the number of samples can further lower down the norm value. This is consistent with what we get in Fig. S1 for off-diagonal elements.
 
There is not an explicit issue of complexity here in our hardware implementation. We simply build a long-enough chip with random phases and, according to our theory\cite{Banchi2017}, the evolution will converge to a Haar random evolution. Now, when is a chip "long-enough"? This is a hard question: in \cite{Banchi2017} we show that this depends on the spectral gap of a Liouvillian operator, and estimating gaps of many-body operators is hard, see\cite{Cubitt2015}. Nonetheless, there are some noise models whose Liouvillean is exactly solvable, like the one considered in \cite{Banchi2017}. The model considered in this experiment is a noisier version (noise on all waveguides vs noise on just a single waveguide), so the required length would be lower than our theory, which can be considered as an upper bound.

In our photonic setting, we see that generally the lattice with a large $\Delta \beta$ amplitude converges faster than those with a small $\Delta \beta$ amplitude, so we can take advantages of our capacity of large-scale photonic lattice and mature $\Delta \beta$ manipulation techniques to make an efficient implementation of Haar randomness. For instance, for a 5×5 lattice with a $\Delta \beta$ amplitude of 0.6$\rm mm^{-1}$, it can well converge to Haar measure at an evolution length within 10cm, which is experimentally practical. 

We further show that the segment length also influences the required evolution length for convergence in Fig. S13. Segment length suggests how frequently we introduce the random $\Delta \beta$ setting, as the value of $\Delta \beta$ will be constant throughout each segment. For instance, 10cm can be 50 segments multiplies a segment length of 2mm, or 20 segments multiplies a segment length of 5mm. We regard the evolution comes to convergence when norm firstly drops below the convergence value, which is around 0.0003 in this case of a fixed number of samples at 100. In Fig. S13c, each dot of convergence length is obtained by averaging 10 groups of setting, each with a fixed number of samples at 100. The error bars in Fig. S13c are the standard deviation of the convergence length for that 10 groups.

In Fig. S13 we see that while varying the segment length would not alter the norm value at convergence, it does change the convergence length, which can reach a minimal value when choosing a proper segment length at around 5-10mm. In the main experiment we use a $\Delta \beta$ of 0.4$\rm mm^{-1}$ and a segment length $\Delta z$ of 2mm. If still using a $\Delta \beta$ of 0.4$\rm mm^{-1}$, but with a $\Delta z$ of 5mm instead 2mm, it can converge at a faster speed, within a feasible chip length of 9.5cm. Changing a segment length is not of any experimental challenge, especially when enlarging it, so even a $\Delta \beta$ of 0.4$\rm mm^{-1}$ can converge at a reasonably short chip length with properly selected segment length.

\subsection{Supplementary Note 8: Convergence for two or multiple photons}
For two travelling particles injected in modes $i$ and $i'\neq i$ we get at the end of the chip the following state
\begin{equation}
	\ket{\psi(U)} = \sum_{kk'} U_{ki} U_{k'i'} a_{k}^\dagger a_{k'}^\dagger\ket 0.
\end{equation}
Setting $\rho(U)=\ket{\psi(U)}\!\bra{\psi(U)}$ and $\rho=\rho(I)$, where $I$ is the identity matrix, 
the $\ell_2$ norm becomes 
\begin{align}
	\mathcal N_2 =& \left\|\mathbb E_{s}[\rho(U_s)] - \mathbb E_{U\sim\rm Haar}[\rho(U)]\right\|_2^2  \nonumber
							\\=& \mathbb E_{ss'} \Tr[\rho(U_s)\rho(U_{s'})] + \mathbb E_{U,V\sim\rm Haar} \Tr[\rho(U)\rho(V)] -\\&- 
	2\mathbb E_{s}\mathbb E_{U\sim\rm Haar}\Tr[\rho(U_s)\rho(U)]\nonumber 
							\\=& \mathbb E_{ss'} \Tr[\rho(U_{s'}^\dagger U_s)\rho)] + \mathbb E_{U,V\sim\rm Haar} \Tr[\rho(V^\dagger U)\rho] -\\&- 
	2\mathbb E_{s}\mathbb E_{U\sim\rm Haar}\Tr[\rho(U^\dagger U_s)\rho]  \nonumber
							\\=& \mathbb E_{ss'} \Tr[\rho(U_{s'}^\dagger U_s)\rho)] - \mathbb E_{U\sim\rm Haar} \Tr[\rho(U)\rho] 
							\label{N2},
\end{align}
where in the last line we used the property that $UV$ and $VU$ are Haar distributed if
$U$ is Haar distributed and $V$ is any unitary matrix. 
To evaluate the norm we note that 
\begin{align}
	\label{asd1}
	\Tr[\rho(U)\rho] &= |\bra{\psi(I)}\psi(U)\rangle|^2  \\&= \left|\sum_{kk} U_{ki}U_{k'i'} (\delta_{ki}\delta_{k'i'}+\delta_{ki'}\delta_{k'i})\right|^2
	\nonumber
\\ 
									 &=	 \left| U_{ii}U_{i'i'}+ U_{ii'}U_{i'i} \right|^2.
	\nonumber
\end{align}
By explicit computation using results from \cite{Richard2019}
we find $\mathbb E_{\rm Haar}[\rho(U)\rho] =\frac{2}{N(N+1)}$. Inserting the above results in Eq.~\eqref{N2} we get
\begin{align}
	\mathcal N_2 =& \frac1{m^2}\sum_{s,s'} 
	\left| (U_{s'}^\dagger U_s)_{ii}(U_{s'}^\dagger U_s)_{i'i'}+ (U_{s'}^\dagger
	U_s)_{ii'}(U_{s'}^\dagger U_s)_{i'i} \right|^2 - \frac2{N(N+1)}.
\end{align}
As with the single photon experiment, we now focus on measurable quantities, with a similar meaning of the above 
formal norm. 
By a similar calculation to \eqref{asd1} we find that the probability that the two photons 
come out from modes $j$ and $j'$, when they were injected in $i$ and $i'\neq i$ is 
\begin{equation}
	p_{j,j'}(U,i,i') = \frac{\left| U_{ji}U_{j'i'}+ U_{ji'}U_{j'i} \right|^2}{1+\delta_{j,j'}},
\end{equation}
and integrating $U$ 
\begin{align}
	p_{j,j'}^{\rm Haar} = \mathbb U_{\rm Haar} [p_{j,j'}(U,i,i')] = \frac2{N(N+1)}.
\label{p}
\end{align}
The figure of merit for experiments is therefore 
\begin{equation}
	\| \mathbb E_{s} [p(U_s,i,i')] - p^{\rm Haar}\|_2^2 = 
	\sum_{j,j'=1}^N \left(\frac 1m \sum_{s=1}^m p_{j,j'}(U_s,i,i') - p_{j,j'}^{\rm Haar} \right)^2,
\end{equation}
where $p_{j,j'}(U_s,i,i') $ can be measured in experiments with sample $s$ when two photons are injected 
in modes $i$ and $i'$ and detected in modes $j$ and $j'$. 
Another measurable quantity is the intensity on mode $j$
\begin{equation}
	I_{j} (U,i,i') = 2p_{j,j}(U,i,i')+\sum_{j' \neq j} p_{j,j'}(U,i,i').
\end{equation}
Using Eq.~\eqref{p} we find a uniform average intensity on an modes
\begin{equation}
	I_{j}^{Haar}=\frac{2}{N}.
\end{equation}

We now numerically investigate the convergence when injecting two indistinguishable photons. We denote $M_2=\mathbb E_{s}[I_{j}]-\frac{2}{N}$, which is consistent with $\mathbb E_{s}[p(U_s,i,i')]-\frac2{N(N+1)}$. According to the Haar measure, the norm of $M_2$ would converge to zero after a certain evolution length. As shown in Fig.S16, there is indeed a convergence to an ultra small value, which is very similar to the single photon scenario presented in Fig.S12. 

With a closer comparison with Fig.S12, we can clearly see that the convergence follows a same dependence on $\Delta \beta$ amplitude and number of groups as with the scenario in $q=1$. That is, increasing the $\Delta \beta$ amplitude can speed up the convergence without altering the norm lower bound, while increasing the number of samples can further lower down the norm. Another similarity is that, both $q=2$ and $q=1$ converge at the similar evolution length, e.g. around 30 cm using a $\Delta \beta$ amplitude of 0.3${\rm mm}^{-1}$. A small difference lies in that the lower bound for $q=2$ is slightly higher, e.g., 0.0008 for $q=1$ and 0.001 for $q=2$ with 500 samples. In fact, the lower bound is a less relevant quantity, as it can slightly compensated by increasing the number of samples, and meanwhile, if we want to run a boson sampling experiment we will always use a single sample. 

We suggest, from an analytical point of view, for the spectral norm we predict no dependence on $q$, because the spectral norm depends only on the largest singular value, not on the dimension of the operators. The numerical analysis of the experimental data uses on the other hand the Frobenius norm. The latter depends on all singular values and hence may scale with the dimension of the Hilbert space, which is larger for more photons. We suggest this scaling factor should be linear in the dimension, so not too big for the current experiment. Just as what we have numerically shown, with 500 samples, $q=1$ reaches 0.0008 and $q=2$ reaches 0.001. In order to make $q=2$ reach 0.0008, we may use a few more samples. As can be inferred by comparing Fig.16a and b, adding the sample number would slightly enhance the required evolution length, only by a very marginal value. Therefore, we suggest this experimental approach for the convergence to Haar measure is scalable for those beyond single photon scenarios.   

More generally we may write the intensity as
\begin{equation}
	I_{j}(U) = \langle \psi (U)|a_{j}^{\dagger} a_{j}|\psi (U)\rangle = \sum_{kk'} U_{k,j}^{*}U_{k',i}\langle \psi |a_{k}^{\dagger} a_{k'}|\psi \rangle,
\end{equation}
so there is:
\begin{equation}
	I_{j}^{Haar} (U) =\sum_{k} \frac{1}{N}\langle \psi (U) |a_{k}^{\dagger} a_{k}|\psi (U) \rangle = \frac{q}{N},
\end{equation}
where $q$ is the number of photons.
As for the uniform probabilities we may use Theorem 4.3 in \cite{Richard2019} to write 
\begin{equation}
\begin{split}
	p_{j_{1},j_{2},…,j_{q}}^{Haar} =\frac{1}{C_{N+q-1}^{q}} \langle j_{1},…,j_{q}|P_{s}|j_{1},…,j_{q}\rangle \\ = \frac{1}{C_{N+q-1}^{q}} \equiv \prod_{k=1}^{q} \frac{k}{N+q-k},
\end{split}
\end{equation}
where $P_{s}$ is the projection onto the symmetric subspace, and in the second line we used the fact that multiphoton Fock state are normalized and symmetric. For $q=1$ we get the usual $1/N$, while for $q=2$ we recover Eq.~\eqref{p}.

\newpage
\begin{figure*}[]
\includegraphics[width=0.46\textwidth]{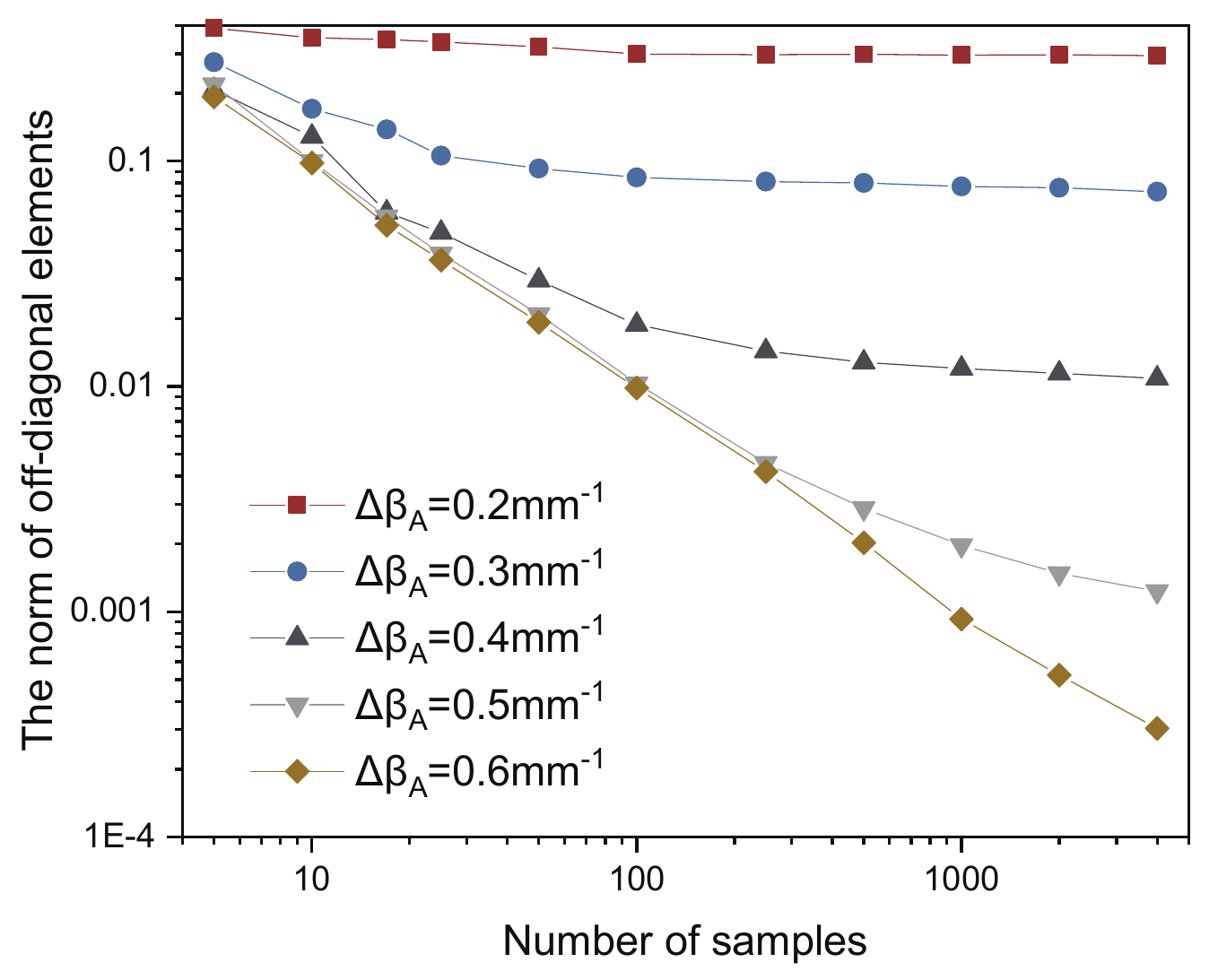}
\caption{\textbf{Convergence of the off-diagonal elements when the number of samples increases.} The numerical L2 norm for the off-diagnal elements of $\mathbb{E}_i[U_{i}\rho U_{i}^{\dagger}]-I/N$ for different numbers of samples at an evolution length of 10cm. $U$ is just calculated as $U(z)=(\prod_k e^{-iH_{\rm eff}(k)\Delta z})$ and $H_{\rm eff}$ follows the $\Delta \beta$ approach given in Eq.(3) of the main text. The curves for QSW corresponds to a $\Delta \beta$ amplitude, $\Delta \beta_A$, of 0.2-0.6 ${\rm mm}^{-1}$, respectively. For all these samples, the segment length $\Delta z$ is 2mm.}  
\label{fig:apparato}
\end{figure*}

\begin{figure*}[]
\includegraphics[width=0.75\textwidth]{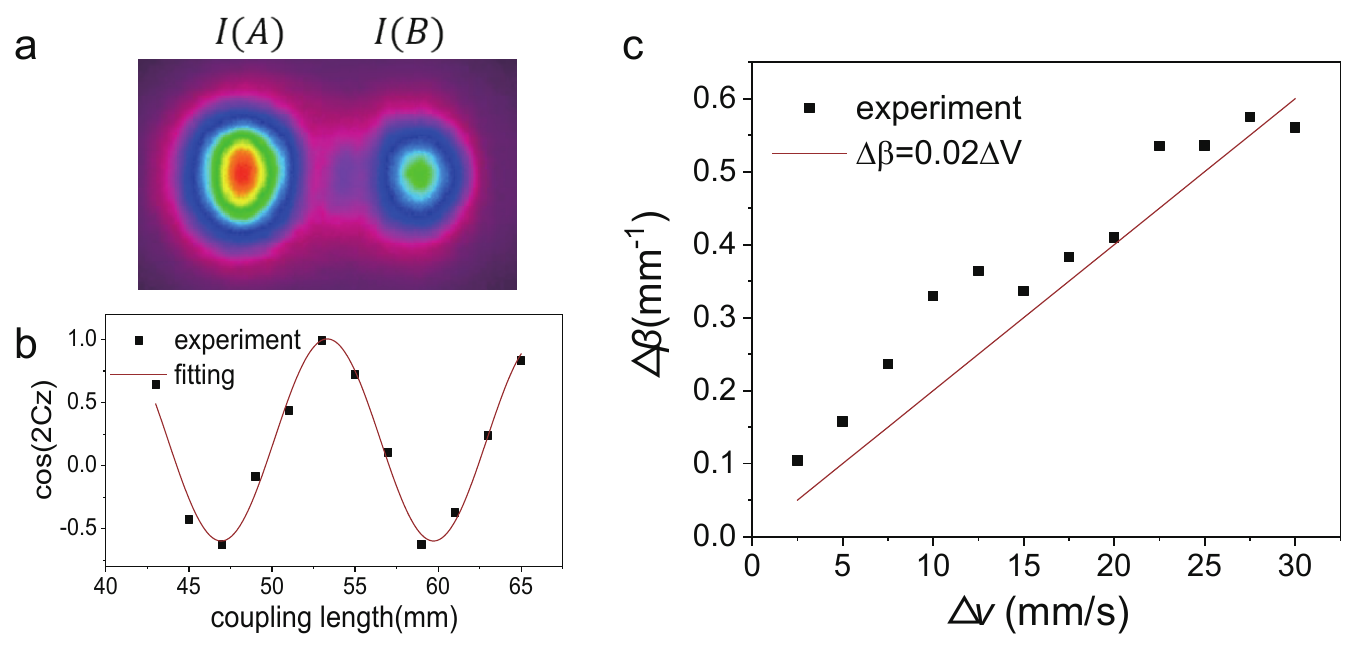}
\caption{\textbf{The characterization of the $\Delta \beta$.} (a) The measurement of the intensity of the two modes, I(A) and I(B), of the directional coupler of a certain evolution length. (b) Fit the relationship between $\frac{I(A)-I(B)}{I(A)+I(B)}$ and $z$ with a suitable sinusoidal curve to get the coupling coefficient. (c) The experimentally measured $\Delta \beta$ against $\Delta V$.}
\label{fig:apparato}
\end{figure*}

\begin{figure*}[]
\includegraphics[width=0.85\textwidth]{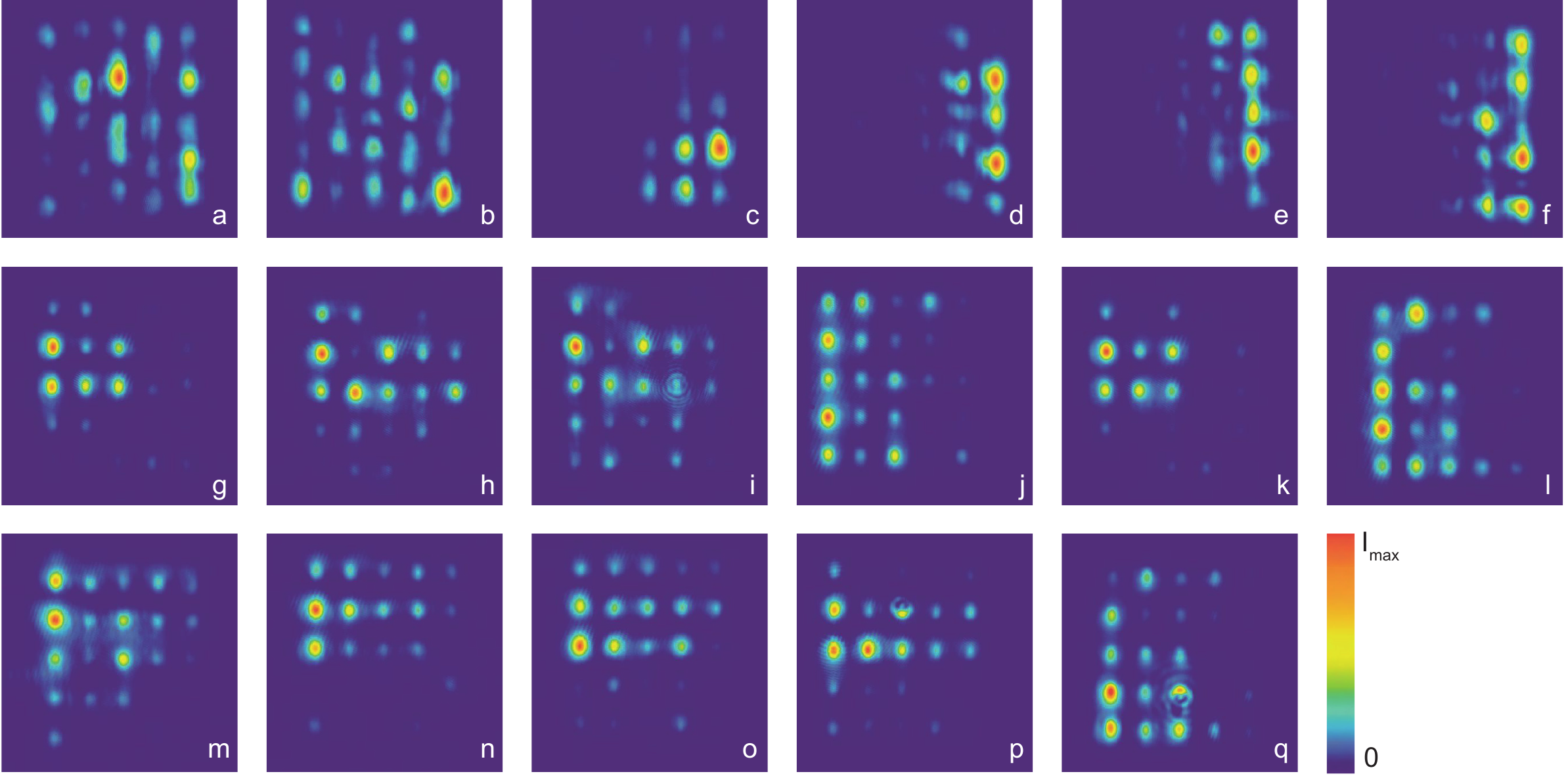}
\caption{\textbf{The 2D patterns for an evolution length of 1cm.} (a-q) The experimentally measured evolution patterns with 17 different random settings. For all these samples, the $\Delta \beta$ amplitude is 0.4 ${\rm mm}^{-1}$ and the segment length $\Delta z$ is 2mm. }
\label{fig:apparato}
\end{figure*}

\begin{figure*}[]
\includegraphics[width=0.85\textwidth]{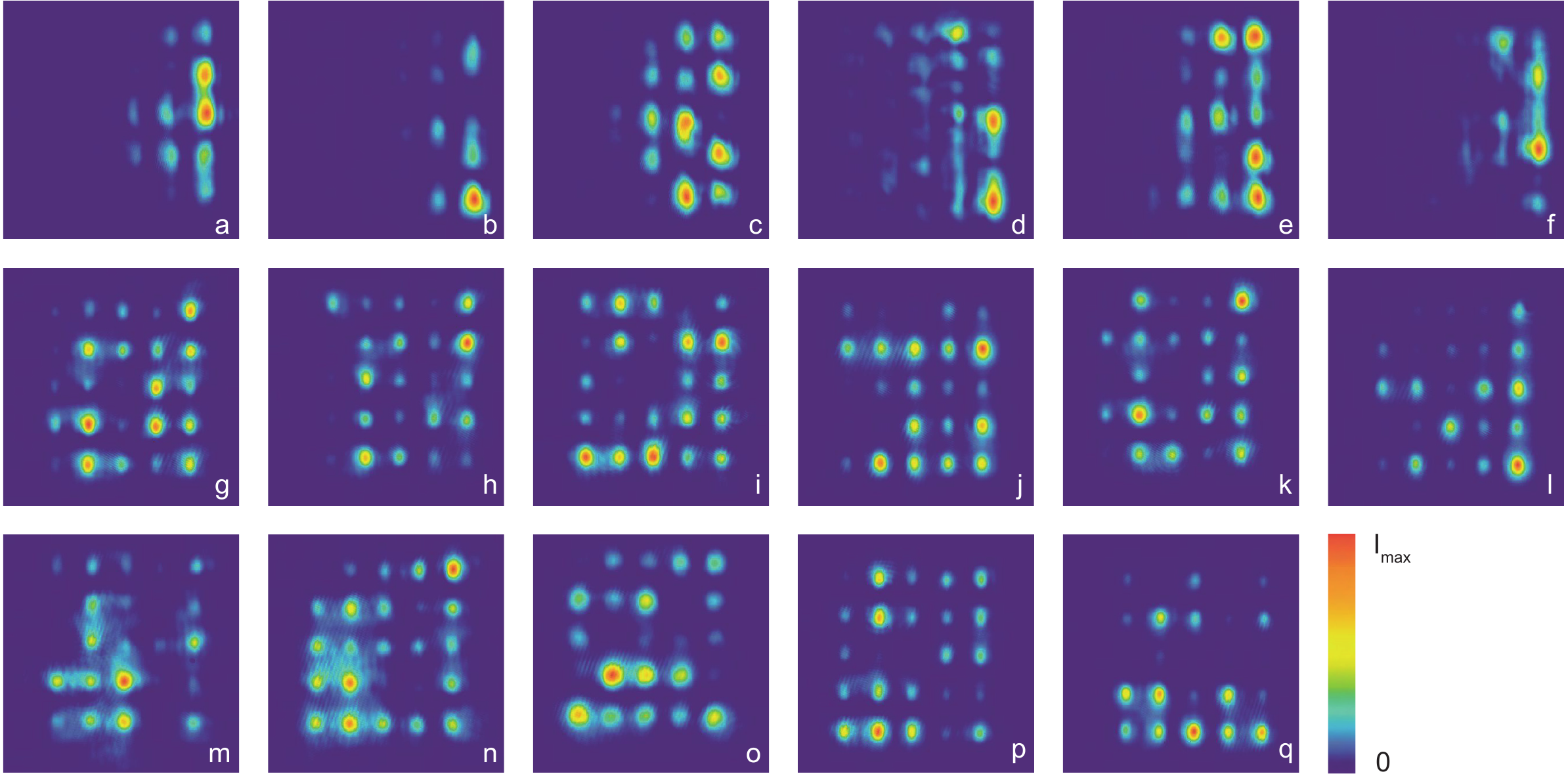}
\caption{\textbf{The 2D patterns for an evolution length of 2cm.} (a-q) The experimentally measured evolution patterns with 17 different random settings. For all these samples, the $\Delta \beta$ amplitude is 0.4 ${\rm mm}^{-1}$ and the segment length $\Delta z$ is 2mm. }
\label{fig:apparato}
\end{figure*}

\begin{figure*}[]
\includegraphics[width=0.85\textwidth]{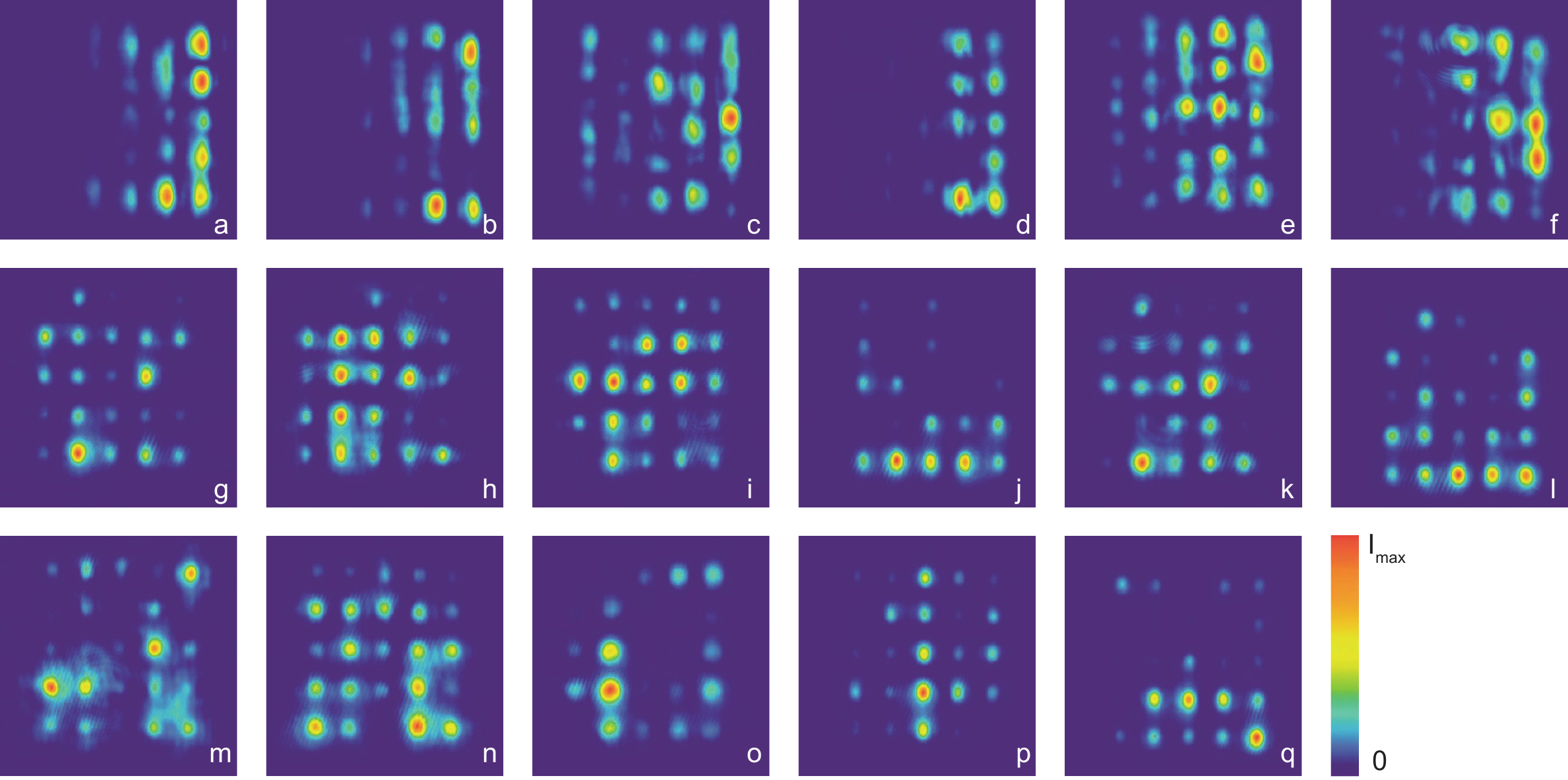}
\caption{\textbf{The 2D patterns for an evolution length of 3cm.} (a-q) The experimentally measured evolution patterns with 17 different random settings. For all these samples, the $\Delta \beta$ amplitude is 0.4 ${\rm mm}^{-1}$ and the segment length $\Delta z$ is 2mm. }
\label{fig:apparato}
\end{figure*}

\begin{figure*}[]
\includegraphics[width=0.85\textwidth]{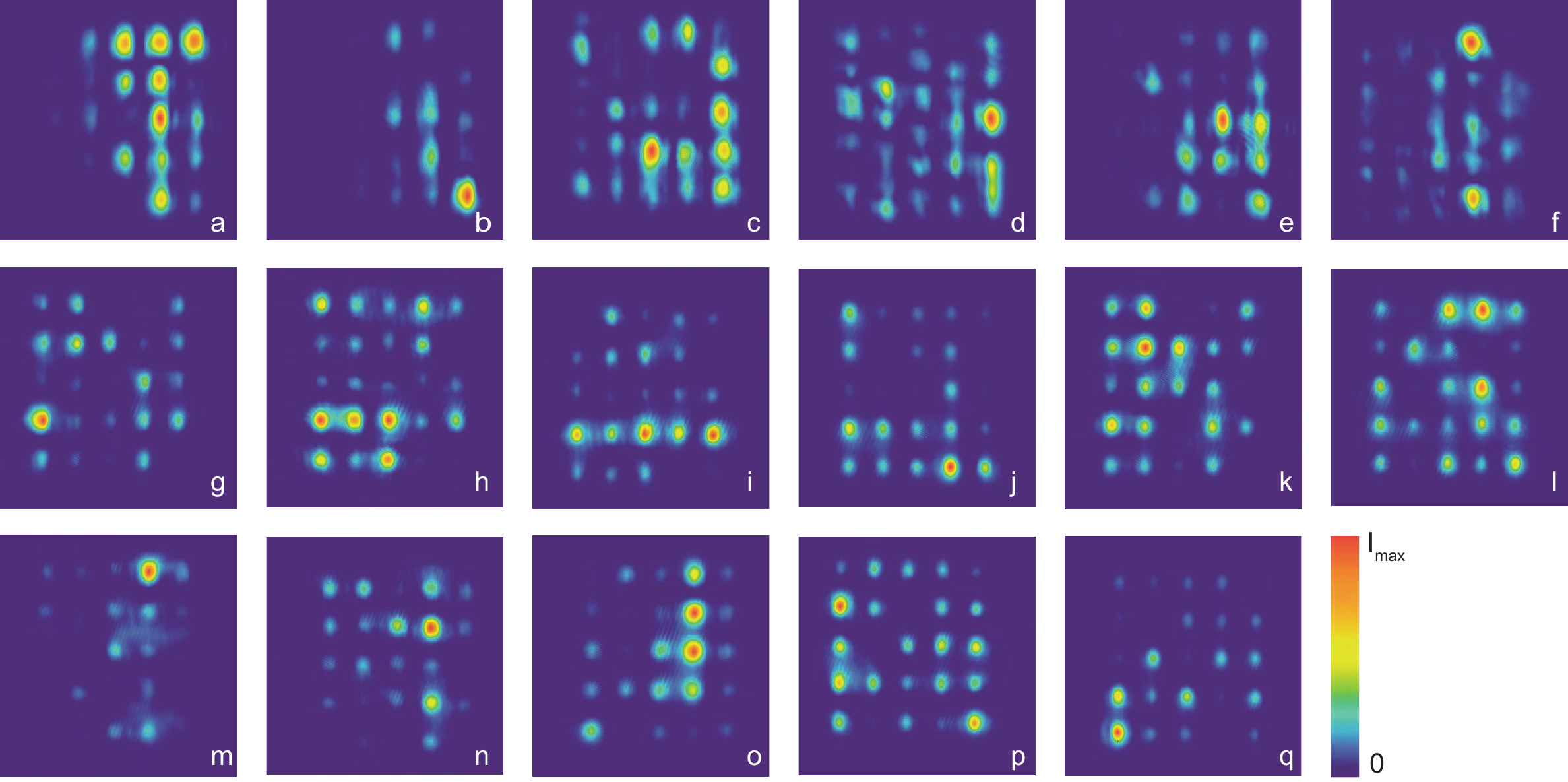}
\caption{\textbf{The 2D patterns for an evolution length of 4cm.} (a-q) The experimentally measured evolution patterns with 17 different random settings. For all these samples, the $\Delta \beta$ amplitude is 0.4 ${\rm mm}^{-1}$ and the segment length $\Delta z$ is 2mm. }
\label{fig:apparato}
\end{figure*}

\begin{figure*}[]
\includegraphics[width=0.85\textwidth]{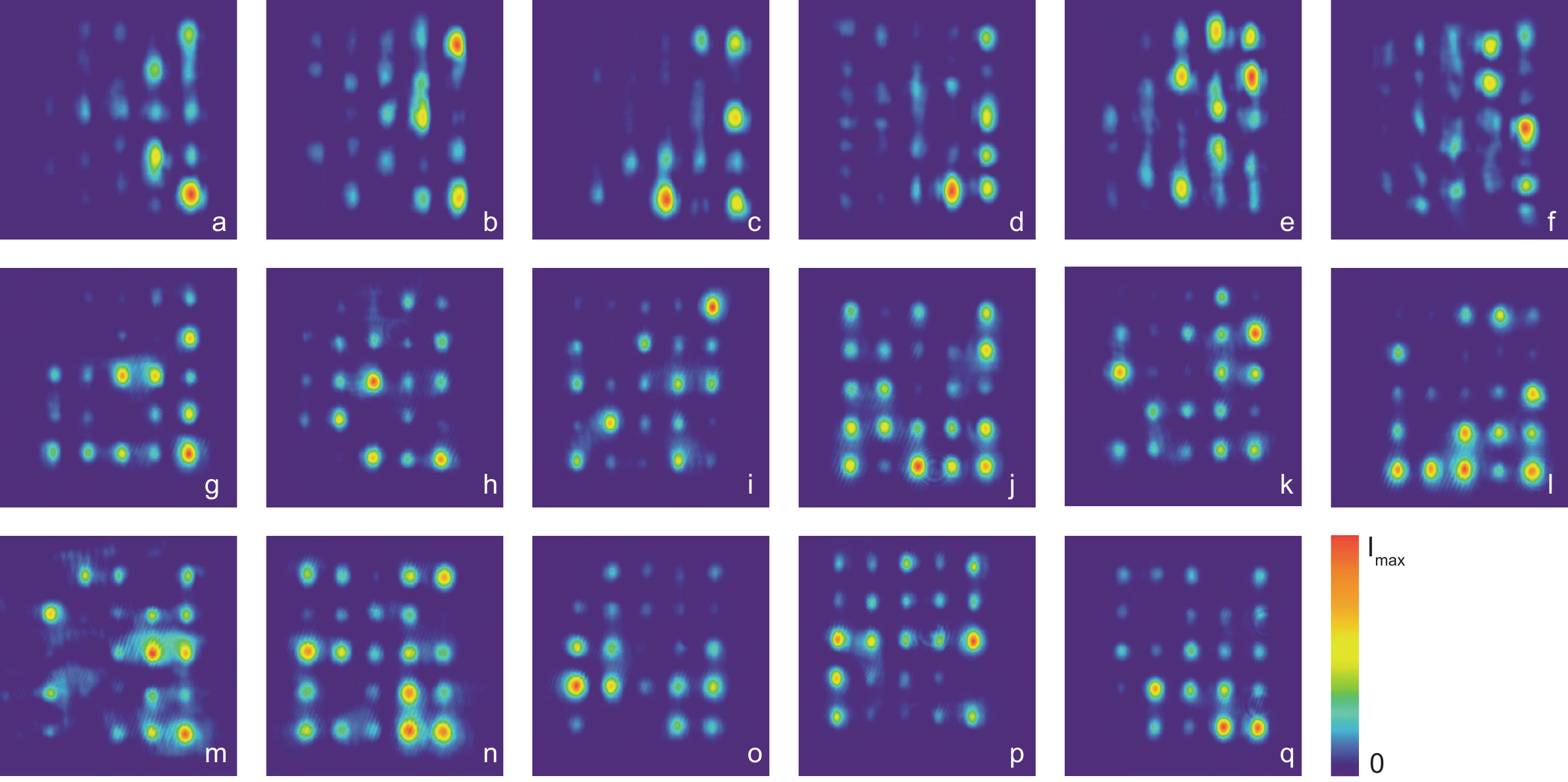}
\caption{\textbf{The 2D patterns for an evolution length of 5cm.} (a-q) The experimentally measured evolution patterns with 17 different random settings. For all these samples, the $\Delta \beta$ amplitude is 0.4 ${\rm mm}^{-1}$ and the segment length $\Delta z$ is 2mm. }
\label{fig:apparato}
\end{figure*}

\begin{figure*}[]
\includegraphics[width=0.85\textwidth]{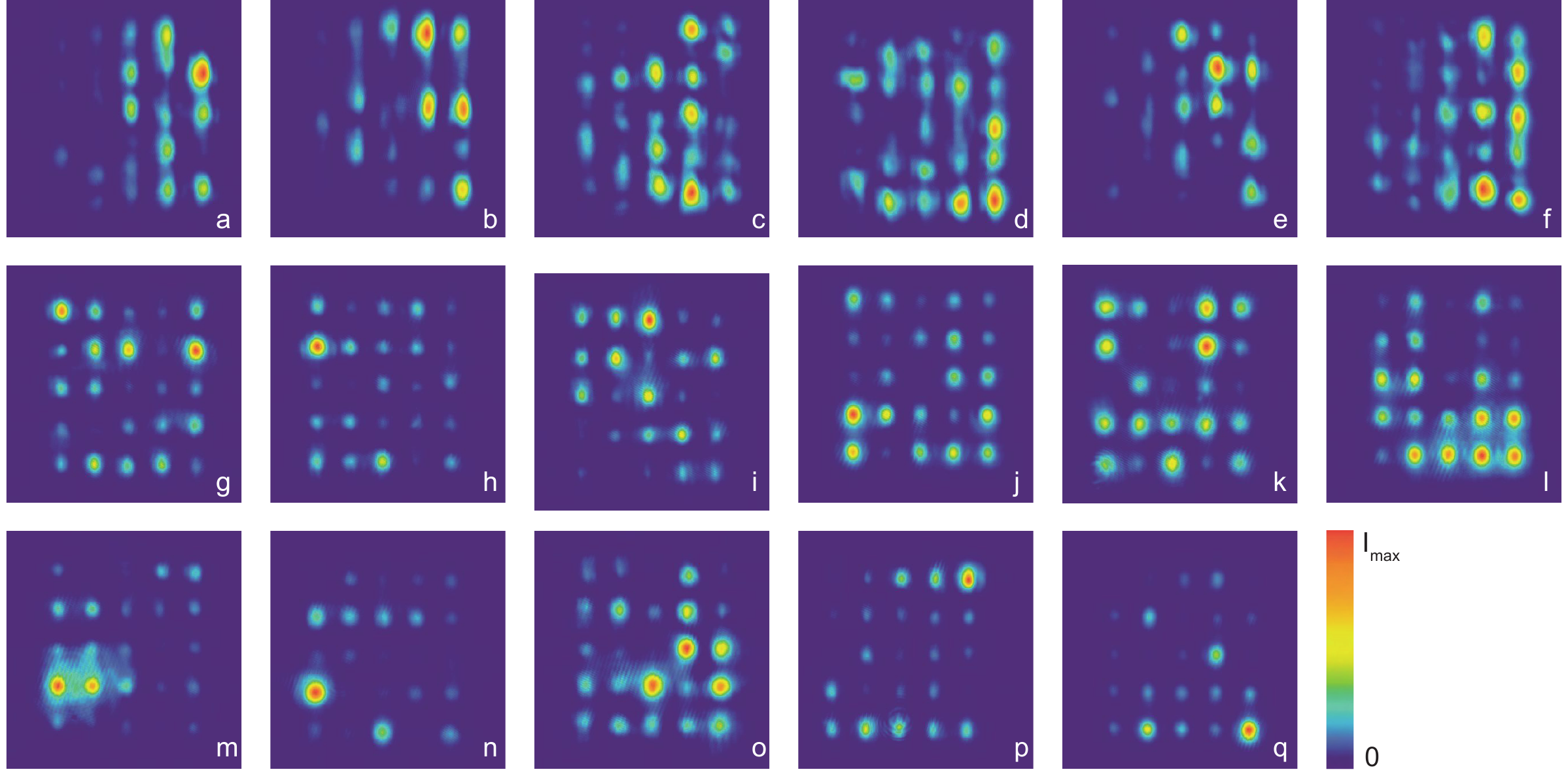}
\caption{\textbf{The 2D patterns for an evolution length of 6cm.} (a-q) The experimentally measured evolution patterns with 17 different random settings. For all these samples, the $\Delta \beta$ amplitude is 0.4 ${\rm mm}^{-1}$ and the segment length $\Delta z$ is 2mm. }
\label{fig:apparato}
\end{figure*}

\begin{figure*}[]
\includegraphics[width=0.85\textwidth]{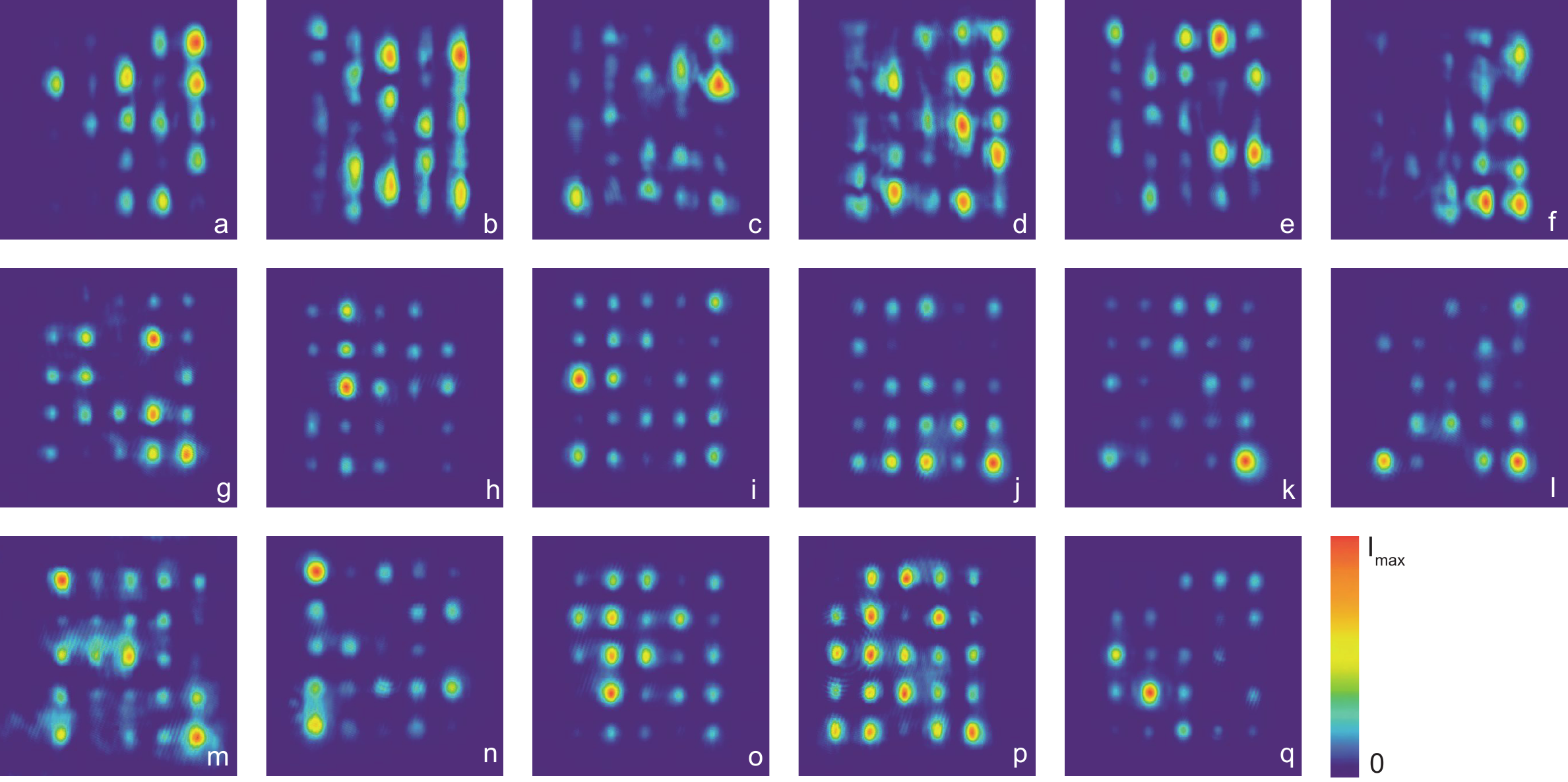}
\caption{\textbf{The 2D patterns for an evolution length of 7cm.} (a-q) The experimentally measured evolution patterns with 17 different random settings. For all these samples, the $\Delta \beta$ amplitude is 0.4 ${\rm mm}^{-1}$ and the segment length $\Delta z$ is 2mm. }
\label{fig:apparato}
\end{figure*}

\begin{figure*}[]
\includegraphics[width=0.85\textwidth]{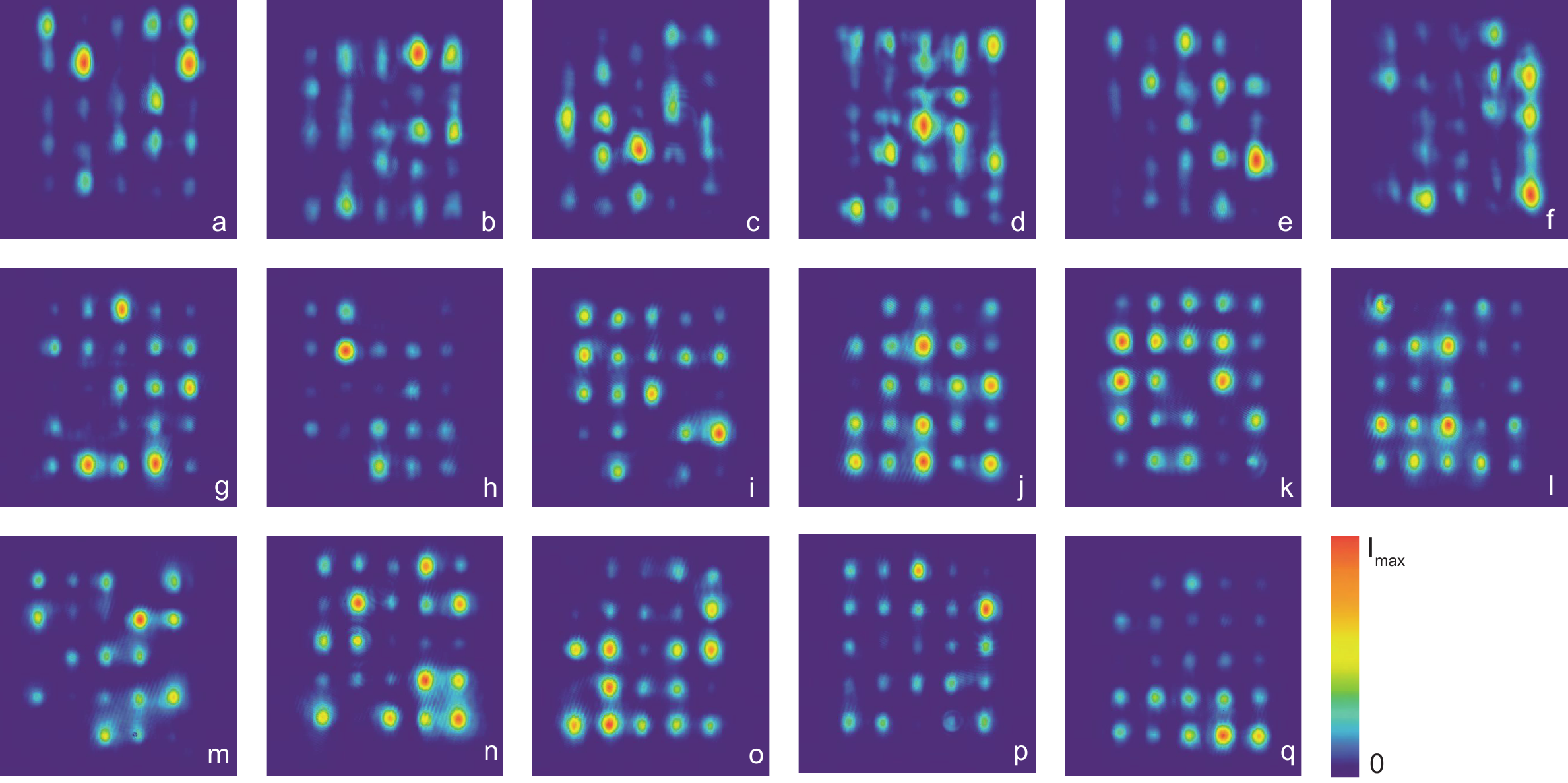}
\caption{\textbf{The 2D patterns for an evolution length of 8cm.} (a-q) The experimentally measured evolution patterns with 17 different random settings. For all these samples, the $\Delta \beta$ amplitude is 0.4 ${\rm mm}^{-1}$ and the segment length $\Delta z$ is 2mm. }
\label{fig:apparato}
\end{figure*}

\begin{figure*}[]
\includegraphics[width=0.46\textwidth]{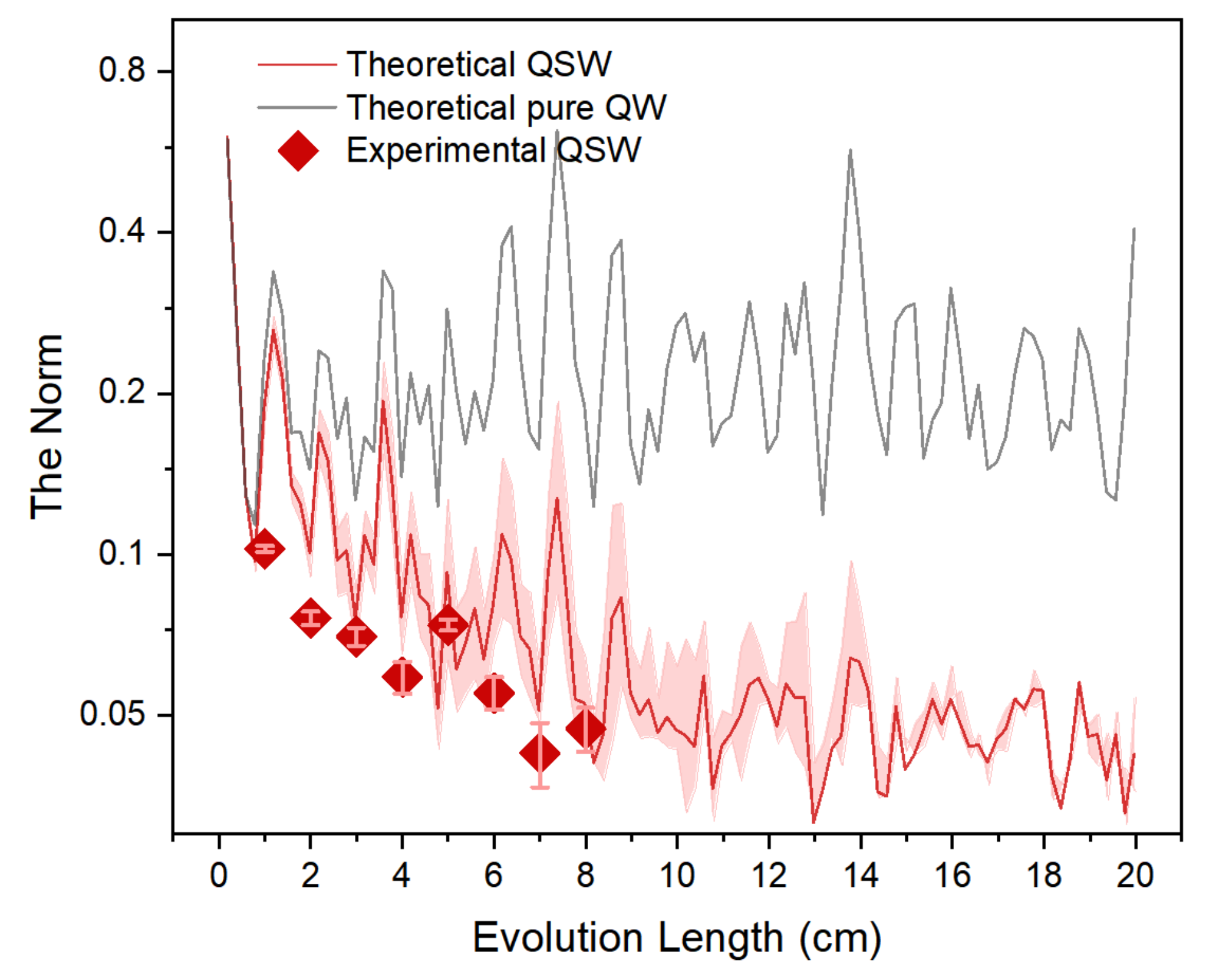}
\caption{\textbf{Convergence of $\Vert M \Vert$.} The L2 norm $\Vert M \Vert$ for 17 samples of different evolution lengths. The theoretical results are obtained using the same setting with that in in Fig. 2. The curve for QSW corresponds to a $\Delta \beta$ amplitude of 0.4${\rm mm}^{-1}$. The shading area shows the possible theoretical result considering a 15$\%$ fluction of the $\Delta \beta$ amplitude, that is, the upper bound and lower bound of the shading area correspond to a $\Delta \beta$ amplitude of 0.34${\rm mm}^{-1}$ and 0.46${\rm mm}^{-1}$, respectively.   } 
\label{fig:apparato}
\end{figure*}

\begin{figure*}[]
\includegraphics[width=0.95\textwidth]{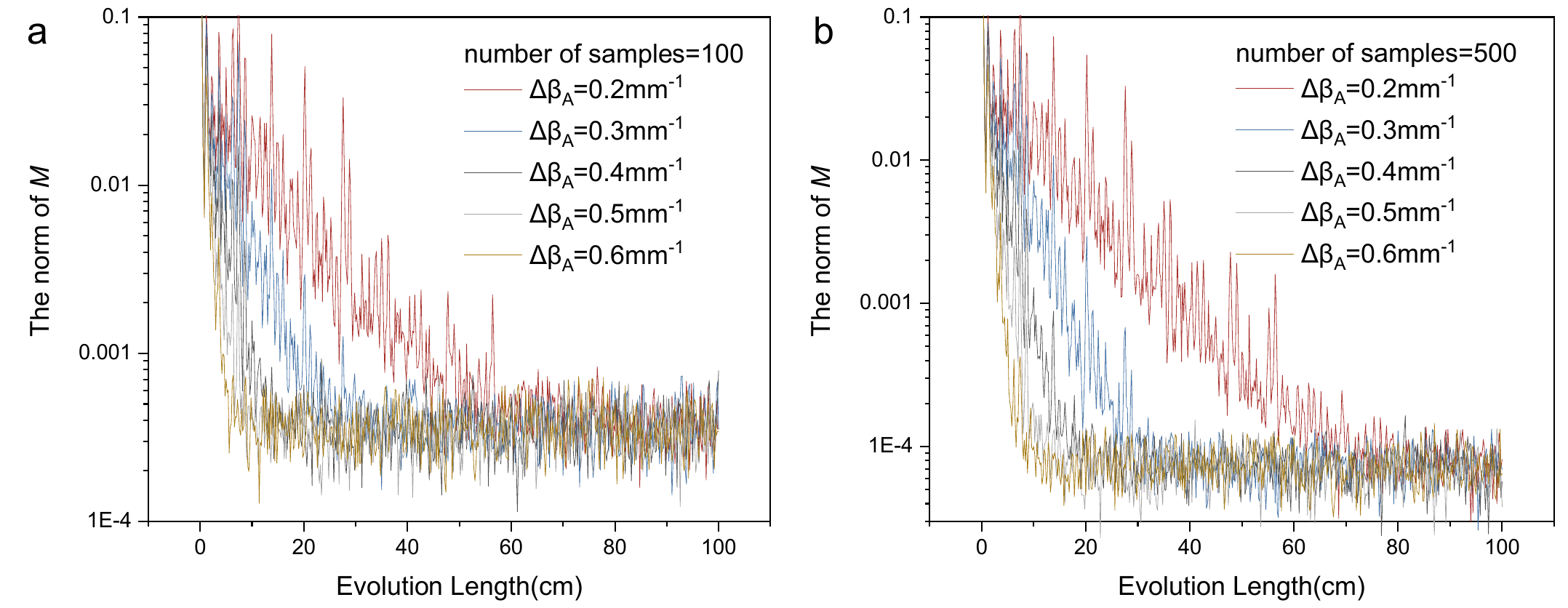}
\caption{\textbf{Convergence of $\Vert M \Vert$ for long evolution lengths.} The numerical L2 norm for the diagnal elements of $\mathbb{E}_i[U_{i}\rho U_{i}^{\dagger}]-I/N$ for different evolution lengths with the number of samples fixed at 100 in (a), and with the number of samples fixed at 500 in (b). The curves for QSW correspond to a $\Delta \beta$ amplitude, $\Delta \beta_A$, of 0.2-0.6 ${\rm mm}^{-1}$, respectively. For all these samples, the segment length $\Delta z$ is 2mm.}
\label{fig:apparato}
\end{figure*}

\begin{figure*}[]
\includegraphics[width=0.99\textwidth]{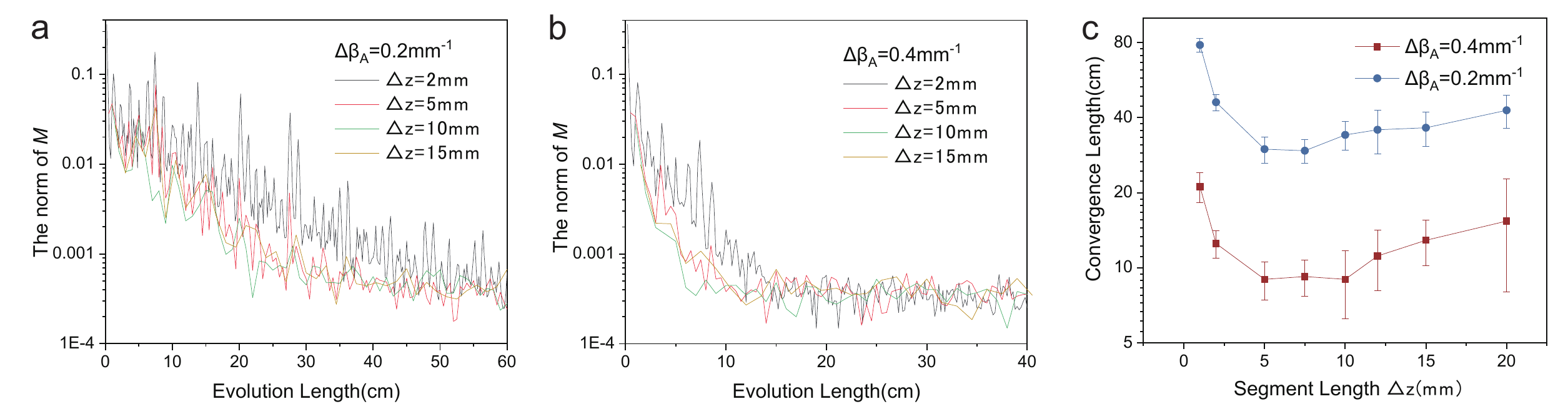}
\caption{\textbf{Convergence with different segment lengths.} The numerical L2 norm for the diagnal elements of $\mathbb{E}_i[U_{i}\rho U_{i}^{\dagger}]-I/N$ for a $\Delta \beta$ amplitude of 0.2 ${\rm mm}^{-1}$ in (a), and for a $\Delta \beta$ amplitude of 0.4 ${\rm mm}^{-1}$ in (b). The curves correspond to different choices of segment lengths $\Delta z$. In both (a) and (b), the number of samples are fixed at 100. (c)The convergence length for a $\Delta \beta$ amplitude of 0.2 ${\rm mm}^{-1}$ or 0.4 ${\rm mm}^{-1}$ at different segment length $\Delta z$. } 
\label{fig:apparato}
\end{figure*}

\begin{figure*}[]
\includegraphics[width=0.95\textwidth]{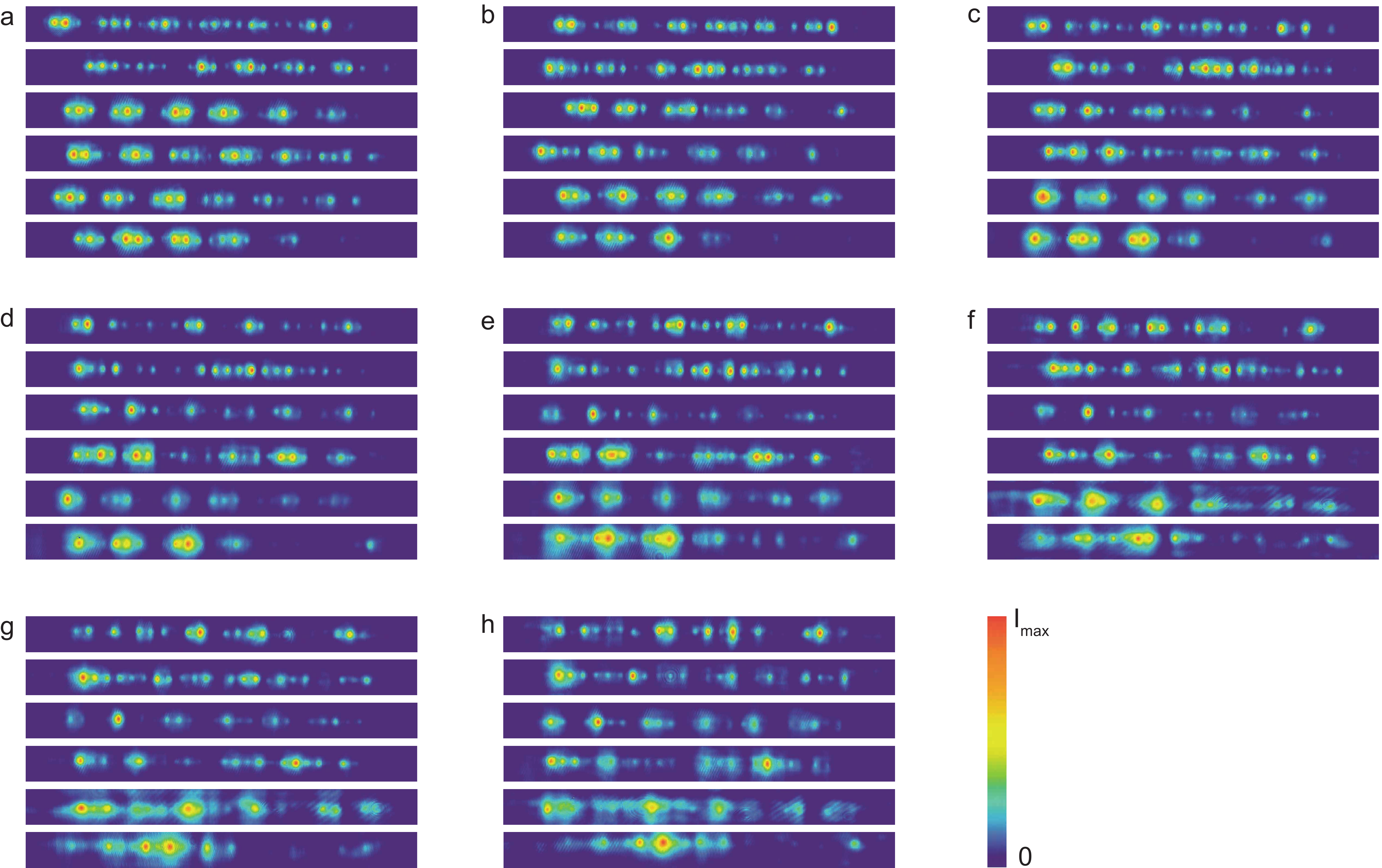}
\caption{\textbf{The 1D patterns for different $\Delta \beta$ amplitudes.} The experimentally measured evolution patterns for samples of a $\Delta \beta$ amplitude of 0.1${\rm mm}^{-1}$ in (a), 0.2${\rm mm}^{-1}$ in (b), 0.3${\rm mm}^{-1}$ in (c), 0.4${\rm mm}^{-1}$ in (d), 0.5${\rm mm}^{-1}$ in (e), 0.6${\rm mm}^{-1}$ in (f), 0.7${\rm mm}^{-1}$ in (g) and 0.8${\rm mm}^{-1}$ in (h), respectively. For each group in (a-h), it includes 6 samples of different random settings. The 6 samples are listed in a column, with 8 columns in total. This can also be viewed as 6 rows. Each row has the same random setting but has 8 different $\Delta \beta$ amplitudes ranging from 0.1${\rm mm}^{-1}$ to 0.8${\rm mm}^{-1}$. For all these samples, the segment length $\Delta z$ is 2mm.}
\label{fig:apparato}
\end{figure*}

\begin{figure*}[]
\includegraphics[width=0.8\textwidth]{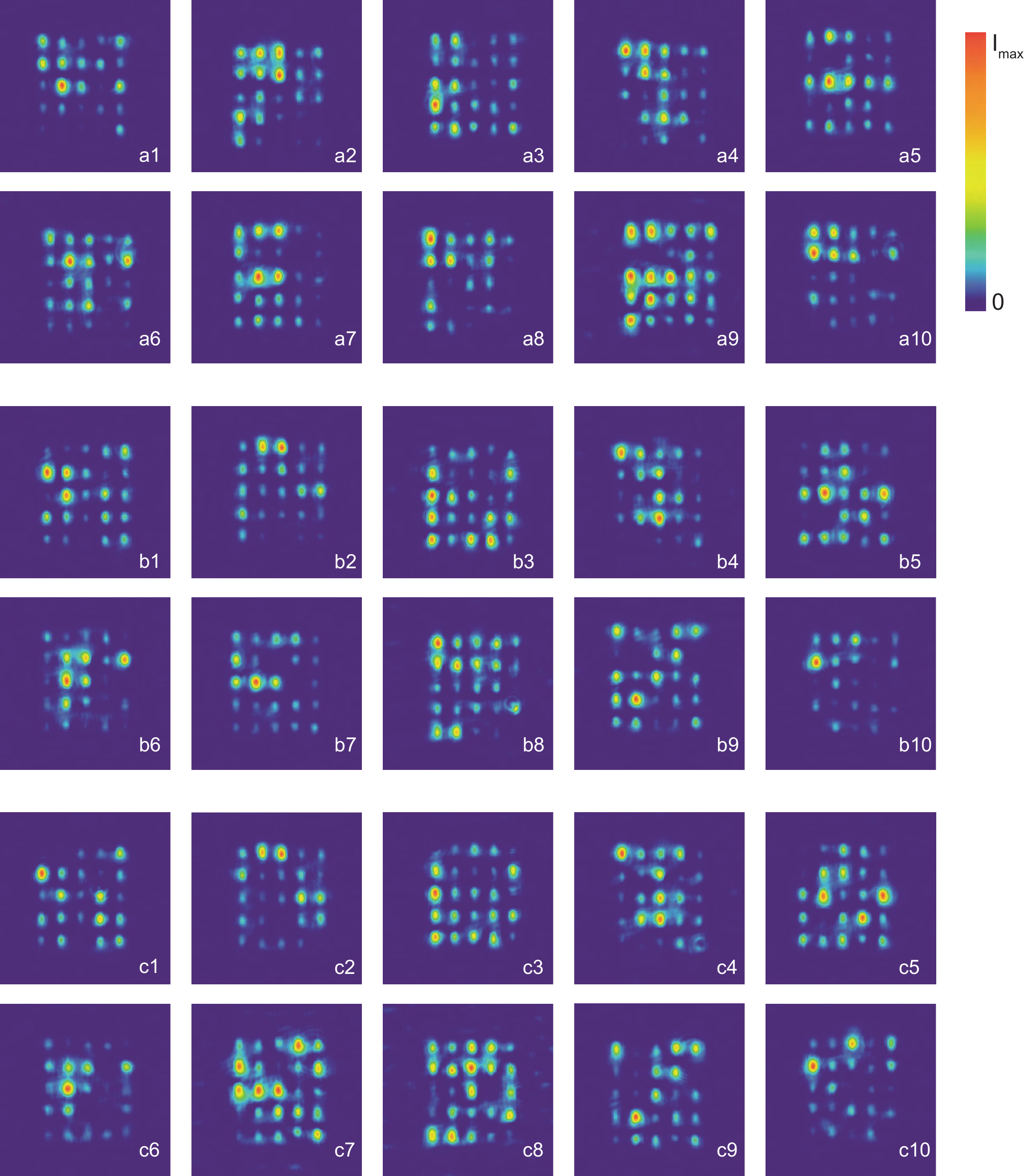}
\caption{\textbf{The 2D patterns for different $\Delta \beta$ amplitudes.} The experimentally measured evolution patterns for samples of a $\Delta \beta$ amplitude of 0.2${\rm mm}^{-1}$ in (a), 0.4${\rm mm}^{-1}$ in (b) and 0.6${\rm mm}^{-1}$ in (c). For each group in (a-c), it includes 10 samples of different random settings. (a1), (b1) and (c1) have the same random setting but different $\Delta \beta$ amplitudes. Same goes for (a2) to (a10). All samples have the same segment length $\Delta z$ of 5mm, and the same evolution length of 5cm. The measured norm for the group with a $\Delta \beta$ amplitude of 0.2${\rm mm}^{-1}$, 0.4${\rm mm}^{-1}$ and 0.6${\rm mm}^{-1}$ are 0.0855, 0.0671 and 0.0534, respectively.}
\label{fig:apparato}
\end{figure*}

\begin{figure*}[]
\includegraphics[width=0.95\textwidth]{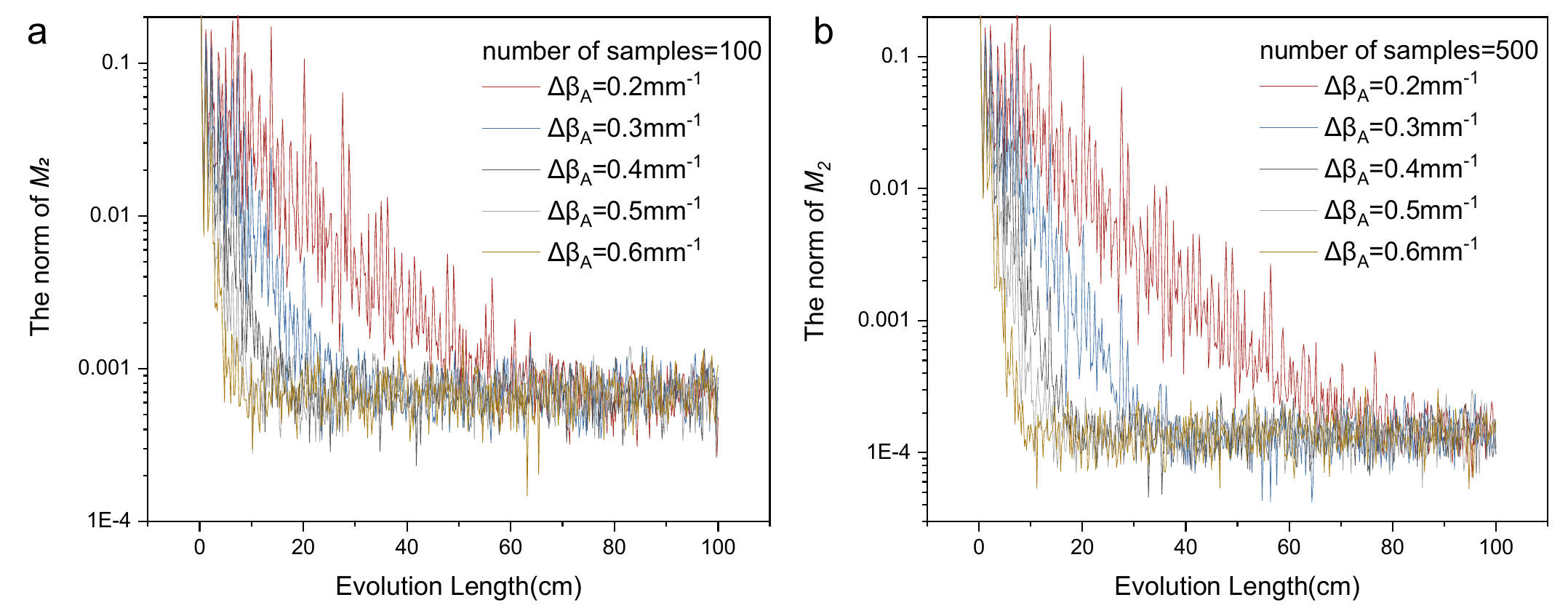}
\caption{\textbf{Convergence of $\Vert M_2 \Vert$ when injecting two indistinguishable photons.} The numerical L2 norm for $M_2=\mathbb E_{s}[I_{j}]-\frac{2}{N}$ for different evolution lengths with the number of samples fixed at 100 in (a), and with the number of samples fixed at 500 in (b). The curves for QSW corresponds to a $\Delta \beta$ amplitude, $\Delta \beta_A$, of 0.2-0.6 ${\rm mm}^{-1}$, respectively. For all these samples, the segment length $\Delta z$ is 2mm.}
\label{fig:apparato}
\end{figure*}

\end{document}